\documentclass[twocolumn]{aastex631}

\usepackage{amsmath}
\usepackage{url}
\usepackage{appendix}
\graphicspath{{./}{figures/}}

\begin{document}

\title{Multi-wavelength Emission from Jets and Magnetically Arrested Disks in Nearby Radio Galaxies: Application to M87}
\shortauthors{Kuze et al.}

\correspondingauthor{Riku Kuze}
\email{r.kuze@astr.tohoku.ac.jp}

\author[0000-0002-5916-788X]{Riku Kuze}
\affiliation{Astronomical Institute, Graduate School of Science, Tohoku University, Sendai 980-8578, Japan}

\author[0000-0003-2579-7266]{Shigeo S. Kimura}
\affiliation{Frontier Research Institute for Interdisciplinary Sciences, Tohoku University, Sendai 980-8578, Japan}
\affiliation{Astronomical Institute, Graduate School of Science, Tohoku University, Sendai 980-8578, Japan}

\author[0000-0002-7114-6010]{Kenji Toma}
\affiliation{Frontier Research Institute for Interdisciplinary Sciences, Tohoku University, Sendai 980-8578, Japan}
\affiliation{Astronomical Institute, Graduate School of Science, Tohoku University, Sendai 980-8578, Japan}



\begin{abstract}
Radio galaxies are a subclass of active galactic nuclei that drive relativistic jets from their center and are observed in radio to very-high-energy gamma rays. The emission mechanisms and regions are still unknown. High-energy gamma rays can be explained by the emission from the magnetically arrested disks (MADs) around the central supermassive black hole, for which the magnetic flux threading the black hole is in a saturation level, although the emission from the MADs does not explain the optical and X-ray data. We construct a two-zone multi-wavelength emission model in which optical and X-rays come from jets, while mm/sub-mm and gamma rays come from MADs. Our model takes into account the particle injection by the magnetic reconnection at the jet base close to the black hole and particle entrainment from the ambient gas at the jet emission zone. We apply our model to M87 and find that our model can explain the simultaneous multi-wavelength data, except for the radio data which could be explained if we extend our one-zone emission model to a one-dimensional one. We also find that the strong plasma entrainment is necessary to explain the multi-wavelength data. Our model will be tested by variability analysis among the multi-wavelength data.
\end{abstract}



\section{Introduction} \label{sec:intro}
Radio galaxies, a subclass of radio-loud active galactic nuclei whose jets misalign to the observers, are detected in radio to GeV-TeV gamma-rays. Radio, optical, and X-ray telescopes resolve extended jet structure \citep{Blandford2019, Hada2019}. 
One may consider that multi-wavelength photons come from jets, but the emission mechanisms and regions are still under debate.

Leptonic jet models, in which energetic electrons radiate the multi-wavelength photons via synchrotron and inverse Compton scattering, are considered as the standard scenario \citep[e.g.,][]{Abdo+2009, MAGIC2020}. However, at least for M87, if we try to reproduce the observational data by the leptonic jet model, the magnetic field strength required to explain the multi-wavelength data ($\sim$ 1-10 mG) is lower than that estimated by the observed core-shift in radio band ($\sim$ 1 G)\citep{Kino2015, Jiang2021}. If we use the core-shift-based magnetic field, the calculated gamma-ray flux is far below the observed gamma-ray flux \citep[e.g.,][]{Lucchini+2019, EHT_MWL2021}. 

\cite{Kimura2020} proposed a multi-wavelength emission model from the magnetically arrested disks \citep[MADs; ][]{Bisnovatyi-Kogan1974, Narayan2012} around the central supermassive black hole (BH), for which the magnetic flux threading the BH is in a saturation level. The MAD model can reproduce the mm/sub-mm and gamma-ray data, but cannot explain the optical and X-ray data \citep{Kimura2020, Kuze+2022}. This motivates us to construct a two-zone emission model in which the mm/sub-mm and gamma-rays come from MADs and the optical and X-rays come from jets.

In our two-zone model, we carefully consider the physical connection between the jet and the MAD, especially the plasma loading process onto the jet. General relativistic magnetohydrodynamic (GRMHD) simulations show that the plasma particles in the accretion flows cannot enter the jet polar region by the centrifugal force and the magnetic field barrier, 
forming the BH magnetosphere \citep[e.g.,][]{Tchekhovskoy2011, Nakamura2018, Porth+2019}. Electrons can be accelerated by spark gaps in the BH magnetosphere, emitting high-energy photons via inverse Compton scattering. These photons can produce the electron-positron pairs via the Breit-Wheeler process ($\gamma +\gamma \rightarrow e^+ + e^-$). However, general relativistic particle-in-cell (PIC) simulations \citep{LevinsonCerutti2018, Chen+2018, Kisaka2020, Crinquand2021, Kin2024} show that this process will not produce the number of pairs required for the observed radio emission. Recently a plausible plasma loading model is proposed by \citet{Kimura+2022} and \citet{Chen+2022}. In their model, magnetic reconnection in the BH magnetosphere accelerates the electrons that emit high-energy photons via synchrotron radiation, which interact with each other, producing electron-positron pairs via the Breit-Wheeler process. These pairs are injected into the jet and are sufficient to explain the radio emission \citep{Kimura+2022} (see Figure \ref{Schematic_image}).

In this paper, we construct a two-zone multi-wavelength emission model that takes into account both jets and MADs, which we call `Jet-MAD model'. We consider the plasma loading process into the jets in the MAD phase that makes the magnetization parameter of the jets depend on the BH mass and accretion rate (see Equation (\ref{sig_std}) and Appendix \ref{base_mag}). In the jet, the magnetization parameter is much higher than unity, and we consider the dissipation mechanism at the dissipation region as magnetic reconnection. We also take into account the plasma entrainment from the ambient gas in that region.

This paper is organized as follows. We describe the Jet-MAD model in Section \ref{sec:method}. In Section \ref{sec:results}, we apply our model to M87 and find that our model can reproduce the multi-wavelength data. We also investigate the cases of weak and no plasma entrainment from the ambient gas in Section \ref{entrainemt}. In Section \ref{sec:discuss}, we examine other model assumptions and discuss the implications of our Jet-MAD model. We present our conclusion in Section \ref{sec:summary}.

\section{Jet-MAD Model} \label{sec:method}
\begin{figure*}[tbp]
    \centering
    \gridline{\fig{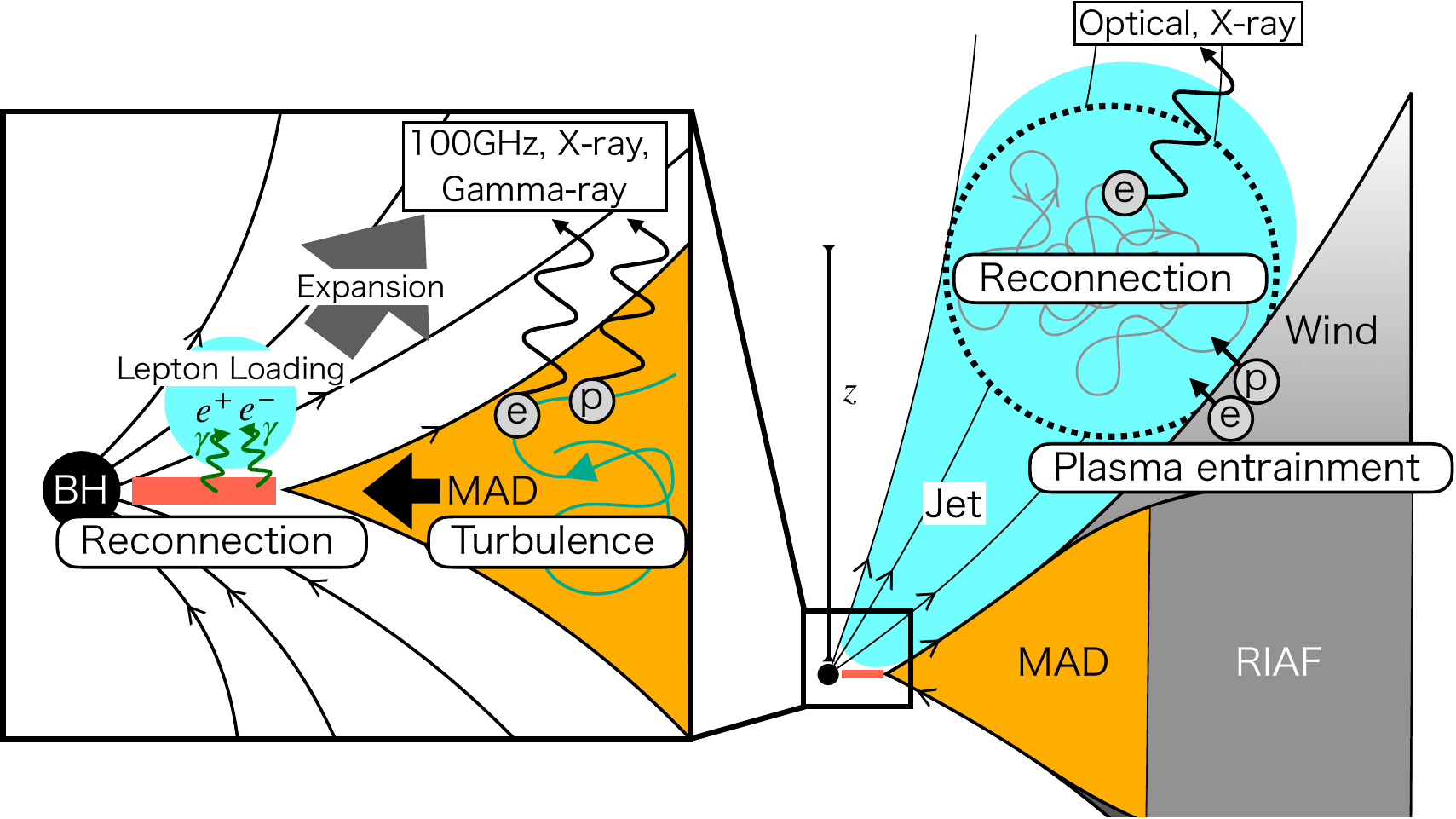}{\linewidth}{}
    }
    \caption{Schematic image of the Jet-MAD model. Left part: Zoom up view around the BH. Protons and electrons inside the MADs emit 100 GHz radio, X-rays, and gamma rays via synchrotron radiation. Magnetic reconnection in the BH magnetosphere accelerates the electrons, and the electrons emit the photons via synchrotron radiation. These photons interact with each other, which form a blob consisting of the electron-positron pairs. This blob expands and becomes the jet material observed in radio galaxies. Right part: Zoom-out view around the jet dissipation region. Magnetic reconnection in the jet accelerates the electrons, which emit optical and X-rays via synchrotron radiation. Plasma is entrained into the jet from the ambient gas.}
    \label{Schematic_image}
\end{figure*}

We construct a multi-wavelength emission model for radio galaxies. In radio galaxies, accretion flows are presumed to be in the MAD regime because of its efficient jet production \citep{Tchekhovskoy2011}, which is also supported by the observations \citep{EHT2021VII, Zamaninasab2014}. We show the schematic image of our Jet-MAD model in Figure \ref{Schematic_image}. In the accretion flows, MHD turbulence is induced by the plasma instability, such as magnetic Rayleigh-Taylor instability \citep[e.g.,][]{Mckinney2012, Marshall+2018, XieZdziarski2019}. The MHD turbulence heats up thermal plasma and accelerates nonthermal particles, leading to multi-wavelength emission as shown in the left part of Figure \ref{Schematic_image}. In the BH magnetosphere, magnetic reconnection is induced near the BH \citep{Ripperda2022}, which accelerates nonthermal electrons efficiently, leading to copious gamma-ray production \citep[e.g.,][]{Hakobyan2023}. This results in the electron-positron pair production, loading a large amount of plasma into the jets \citep{Kimura+2022, Chen+2022}. The loaded plasma expands outward and dissipates its energy via magnetic reconnection, producing multi-wavelength photons via leptonic emissions as shown in the right part of Figure \ref{Schematic_image}. In our Jet-MAD model, the jet and MAD are physically connected via the plasma loading and energy injection processes. In the following subsections, we will explain the individual processes.

\subsection{MAD model} \label{subsec:MAD}
In this subsection, we briefly explain the particle acceleration and emission mechanisms of the MADs. We use the one-zone emission model, called `MAD model', constructed by \cite{Kimura2020} and \cite{Kuze+2022} with modifications of heating and acceleration mechanisms.\footnote{\citet{Kimura2020} and \citet{Kuze+2022} consider the magnetic reconnection as the heating and acceleration processes, while we consider the MHD turbulence dissipation.} We explain the detailed calculation method of the MAD model in Appendix \ref{MAD model}.
We consider that the high-temperature plasma accretes onto the BH of mass $M_{\rm BH}$. The mass accretion rate, $\dot{M}$, and the size of the emission region, $R_d$, are normalized by the Eddington rate and the gravitational radius, respectively, i.e., $\dot{M}c^2  = \dot{m}L_{\rm Edd}$, $R_d = r R_g = r GM_{\rm BH}/c^2$, where $L_{\rm Edd}$ is the Eddington luminosity, $c$ is the speed of light, and $G$ is the gravitational constant. We use the notation $Q_X = Q/10^X$ unless otherwise noted. For the BH mass, we use the notation $M_9 = M_{\rm BH}/(10^9M_\odot)$.

In the MAD model, we consider five particle species: thermal electrons, primary electrons, primary protons, secondary electron-positron pairs produced by the Breit-Wheeler process, and those by the Bethe-Heitler process ($p+\gamma \rightarrow p+e^+ + e^-$). 

 Thermal electrons emit multi-wavelength photons via the synchrotron and Comptonization processes. We determine the thermal electron temperature by equating the electron heating rate with the cooling rate. We consider that the MHD turbulence heats up the thermal electron. Electron heating processes by collisionless MHD turbulence in the MADs are still under debate \citep[cf.][]{Kawazura2020}. Then, we give the ratio of electron-proton heating rate, $Q_i/Q_e$, as a parameter, and we set $Q_i/Q_e \simeq 10$. The cooling rates of the thermal electrons are calculated in the same manner as \citet{Kimura2015}. For M87, the accretion rate is low, and then, the bremsstrahlung is inefficient as a cooling process because of the low electron number density.

MHD turbulence accelerates the primary electrons and protons which emit nonthermal synchrotron radiation \citep{Lynn+2014, Kimura2016, KimuraTomida2019, SunBai2021}. To obtain the energy distributions of the nonthermal particles, we solve the energy transport equation with the one-zone and steady-state approximations, and we use the injection term as a power-law form with an exponential cutoff (see Appendix \ref{MAD model}). We estimate the cutoff energy of the particles by equating the acceleration timescale with the loss timescale. 

As the acceleration process, we consider stochastic acceleration by the MHD turbulence with a hard-sphere-like diffusion coefficient. Within the diffusion process, we write the effective mean free path as $\eta_{\rm turb}H$, where $H$ is the scale height of the accretion flow and $\eta_{\rm turb}$ is the numerical factor. We consider the relevant cooling processes (synchrotron, Bethe-Heitler, photomeson production), diffusive escape, and infall to the BH for the loss processes.

Nonthermal protons interact with photons in radio bands via the Bethe-Heitler process and produce secondary electron-positron pairs. These pairs produce very-high-energy gamma rays, which interact with lower-energy photons in MADs, leading to electromagnetic cascades. We iteratively calculate the photons and the electron-positron pairs spectra until they converge.

\subsection{Physical quantities at the base of the jet} \label{subsec:Jet-MAD}
Recent high-resolution GRMHD simulation shows that magnetic reconnection occurs in the BH magnetosphere \citep{Ripperda2022}. Magnetic reconnection accelerates the electrons, and the accelerated electrons emit high-energy photons. These photons interact with each other via the Breit-Wheeler process and produce electron-positron pairs. These electron-positron pairs are loaded to the jets and lead to the blob formation \citep{Kimura+2022, Chen+2022}.

 In this paper, we estimate the steady-state magnetization parameter in the reconnection region by equating the pair production rate with the pair escape rate \citep[see][]{Chen+2022}. We show the detailed calculation model of the steady-state magnetization parameter in Appendix \ref{base_mag}. According to \citet{Chen+2022}, we estimate the steady-state magnetization parameter by solving 
\begin{equation}
\sigma_\pm = \frac{B_{\rm rec}^2}{4\pi n_\pm m_e c^2},
\label{sig_std}
\end{equation}
 where $\sigma_\pm$ is the magnetization parameter, $B_{\rm rec}$ is the magnetic field in the reconnection region, $n_\pm$ is the produced pair number density in the reconnection region, and $m_e$ is the electron mass.

As $n_\pm$ depends on $\sigma_\pm$ and the size of the reconnection region, $l_{\rm rec}$, the steady-state magnetization parameter depends only on $B_{\rm rec}$ and $l_{\rm rec}$ (see Appendix \ref{base_mag}). We consider that the accretion flow is in the MAD state, and magnetic field strength around the BH is estimated to be $B_H = \sqrt{\dot{M}c\Phi^2_{H}/(4\pi^2R_g^2)}\simeq 1.1\times 10^3 M_9^{-1/2}\dot{m}_{-4}^{1/2}\Phi_{H, 1.7} \ \rm G$, where $\Phi_H \simeq 50 \Phi_{H, 1.7}$ is the saturated magnetic flux of the MAD. High-resolution GRMHD simulation with a BH spin parameter $a = 0.935$ suggests that the magnetic reconnection occurs in a radius $r_{\rm rec} \approx 2 R_g$ \citep{Ripperda2022}, and we fix $r_{\rm rec} = 2 R_g$. We estimate $B_{\rm rec}$ in the same way as \cite{Kimura+2022} and to be $B_{\rm rec} = \sqrt{2} B_H \left( {r_{\rm rec}}/{{R_g}}\right)^{-2}\simeq 3.9\times 10^2M_9^{-1/2}\dot{m}_{-4}^{1/2}\Phi_{\rm rec, 1.2} \ \rm G$, where $\Phi_{\rm rec} = \sqrt{2} \Phi_H (r_{\rm rec}/R_g)^{-2}$ is the magnetic flux in the magnetic reconnection region. We fix $l_{\rm rec}=R_g$ for simplicity.

Finally, we estimate the jet luminosity where the pairs are injected at the jet base as 
\begin{equation}
    L_j = \eta \dot{M}c^2,
    \label{jet_lum}
\end{equation} 
where $\eta$ is the fraction of accretion luminosity to the jet kinetic luminosity. We consider that the energy is injected into the jet by the Blandford-Znajek process \citep{Blandford1977}, and thus, the injected energy flux is proportional to $\sin ^2\theta$, where $\theta$ is the angle between the jet axis to the formed blob. We assume that the blob is spherical and the radius of the blob at the jet base is equivalent to the length of the reconnection region $l_{\rm rec} = R_g$. Then, the angle of the blob to the reconnection region is estimated to be $\tan \theta = l_{\rm rec}/r_{\rm rec} \approx 0.5 \approx \theta $, which leads to $\eta = L_j/L_{\rm BZ} \approx \int^{\pi/2}_{\pi/2 - \theta} \sin^2 \theta d\theta/ (\int^{\pi/2}_{0} \sin^2\theta d\theta) \approx 0.58$, where $L_{\rm BZ}\approx \dot{M}c^2$ is the Blandford-Znajek power. 

\subsection{Energy contents of the jet}\label{energy_content}
In the magnetically dominated jet region, the magnetic field is dominated by the toroidal component and $B \propto R^{-1}$ at $R \gtrsim 10 R_g$, where $R$ is the distance from the jet axis \citep{Kimura+2022}. If all the electron-positron pairs inside the blob are accelerated up to $\sim c$ by efficient electromagnetic bulk acceleration, number flux conservation results in $n_e \propto R^{-2}$. Then, the magnetization parameter is conserved inside the jet, i.e., $\sigma \propto B^2/n_e \propto R^{0}$. However, the bulk acceleration is not efficient as observed in the radio jet of M87 \citep[see Figure 12 of][]{Park2019}. We consider that the single, one-zone blob expands due to the velocity dispersion of electron-positron pairs, along the magnetic field lines, which decreases the number density more rapidly than $n_e \propto R^{-2}$. By employing a numerical factor $\xi$, we write the magnetization parameter at the dissipation region before dissipation as $\xi \sigma_\pm$.

The interval time of the magnetic reconnection flares in the BH magnetosphere is around $\Delta t \simeq 10^3 R_g/c$ \citep{Ripperda2022}, which is comparable to the viscous timescale of the MADs. Thus, we consider that the blob will expand $\sim 10^3$ times along the magnetic field lines.
In this paper, we set $\xi = 10^3$.

We consider that the electron-positron pairs inside the jet are cold due to the jet expansion, and we write the conservation of energy before dissipation as
\begin{align}
    L_j &= \Gamma_j^2\left(n_em_ec^2 + \frac{B^2}{8\pi}\right)\pi R_j^2 \beta_j c \nonumber \\
    &= \Gamma_j^2(1+0.5\xi \sigma_\pm)n_{e}m_ec^2\pi R_j^2 \beta_j c,
\label{eng_consv_bf}
\end{align}
where $R_j$ is the size of the dissipation region, which we treat as a free parameter, $\Gamma_j$ is the jet Lorentz factor, and $\beta_j$ is the jet speed normalized by $c$. We estimate $R_j$ and $\Gamma_j$ in Appendix \ref{Lorentz}. Hereafter, we use the notation $R_{j,2} = R_j /(10^2 R_g)$.

At the dissipation region, the magnetization parameter is $\xi \sigma_\pm \gg 1$, so that converting the magnetic energy to the thermal energy is more efficient than converting the plasma kinetic energy. We consider that the velocity difference between the jet and the ambient gas triggers the Kelvin-Helmholtz instability (KHI) at the edge of the jet \citep{Sironi+2021, Davelaar+2023}, inducing the turbulence. As a result, the electron-proton plasma can be entrained into the jet from the ambient gas, which consists of the wind from the radiatively inefficient accretion flow \citep[RIAF;][]{Narayah&Yi1994, Yuan&Narayan2014} in the outer region of the MAD (see right part of Figure \ref{Schematic_image}). The KHI-induced turbulence also causes magnetic reconnection inside the jet. Thus, the magnetization parameter at the dissipation region decreases due to the plasma entrainment and the dissipation of the magnetic energy through the magnetic reconnection \citep[e.g.,][]{Hoshino2012, Guo2020, Sironi+2021, Mehlhaff2024}. Such turbulent reconnection accelerates nonthermal particles efficiently \citep[e.g.,][]{HoshinoPhyLV2012, Xu2023}. We define the magnetization parameter after the plasma entrainment $\sigma_{\rm ent}$ as a free parameter. The parameter, $\sigma_{\rm ent}$, determines the amount of the plasma entrainment. If the plasma entrainment is strong(weak), it results in low(high) $\sigma_{\rm ent}$. We also define the magnetization parameter after dissipation as $\sigma_{\rm dis}$.

We assume that the entrained electron-proton plasma is cold, and then, we write the conservation of energy after plasma entrainment as 
\begin{align}
    L_j &= \Gamma_j^2\left(n_em_ec^2+n_pm_pc^2+\frac{B^2}{8\pi}\right)\pi R_j^2 \beta_j c \nonumber \\
    &= \Gamma_j^2(1+0.5\sigma_{\rm ent})(n_em_ec^2 + n_pm_pc^2) \pi R_j^2 \beta_j c,
    \label{energy_consv}
\end{align}
where $n_p$ is the proton number density and $m_p$ is the proton mass. If the plasma entrainment is strong ($n_e \approx n_p$), the sum of the electron and proton energy density is written as $n_e m_e c^2+ n_pm_p c^2 \sim n_p m_p c^2$.

Magnetic reconnection dissipates its energy of $\delta B$, where $\delta B$ is the amplitude of the turbulent magnetic field, and the ordered magnetic field is conserved. This implies that the magnetic energy does not change much as long as $\delta B/B < 1$. Also, the magnetic reconnection accelerates the entrained protons, and we estimate the proton thermal energy as $(\delta B/B)^2 \sigma_{\rm ent}m_pc^2$. If $(\delta B/B)^2 \sigma_{\rm ent} < 1$, the proton energy density does not change significantly, and we have $\sigma_{\rm dis} \sim \sigma_{\rm ent}$. In contrast, if $(\delta B/B)^2 \sigma_{\rm ent} > 1$, one has $U_p \approx (\delta B/B)^2 \sigma_{\rm ent}n_pm_pc^2$, and $\sigma_{\rm dis}$ is estimated to be $\sigma_{\rm dis} \approx B^2/(4\pi U_p) \approx (B/\delta B)^2$. As a result, we estimate $\sigma_{\rm dis}$ as
\begin{equation}
    \sigma_{\rm dis} \approx
\begin{cases}
\sigma_{\rm ent} ~~\left( \left(\frac{\delta B}{B}\right)^2 \sigma_{\rm ent} \leq 1 \right) \\
\left(\frac{B}{\delta B}\right)^2  ~~\left( \left(\frac{\delta B}{B}\right)^2 \sigma_{\rm ent} > 1 \right)
\end{cases}.
\label{sig_dis}
\end{equation}

We write the conservation of energy after dissipation as
\begin{align}
      L_j &= L_e+L_p+ \Gamma_j^2\frac{B_{\rm dis}^2}{8\pi} \pi R_j^2\beta_j c \nonumber \\ 
      &= (1+0.5\sigma_{\rm dis})(L_e+L_p),
    \label{after_dis}
\end{align}
where $L_e$ is the electron luminosity, $L_p$ is the proton luminosity, and $B_{\rm dis}$ is the magnetic field after dissipation.

\subsection{Particle distribution inside the jet} \label{distribution}
In this subsection, we show our model to estimate the electron energy distribution in the jet. We only consider the emission from the nonthermal electrons since the emission from the protons is inefficient. To obtain the energy distribution of nonthermal electrons, we solve the transport equation by assuming the one-zone and steady-state approximations:
\begin{equation}
    -\frac{d}{d\gamma} \left(\frac{\gamma N_e(\gamma)}{t_{\rm cool}} \right) = \dot{N}_{e, \rm inj} - \frac{N_e(\gamma)}{t_{\rm esc}},
    \label{eq_transport}
\end{equation}
where $N_e(\gamma)$ is the differential number spectrum, $t_{\rm cool}$ is the cooling time, $t_{\rm esc}$ is the escape time, and $\dot{N}_{e, \rm inj}$ is the injection term. The analytic solution is given by Equation (C.11) in \cite{Dermer2009}. 

We consider that the electron energy distribution accelerated at a certain $\sigma$ is power-law with a hard index and a cutoff Lorentz factor $\gamma_{\rm cut}\sim \sigma (\delta B/B)^2$, and the injection of accelerated electron energy distributions continues from $\sigma = \xi\sigma_\pm$ to $\sigma = \sigma_{\rm ent}$. Hence, we give the superposition of the injected electron energy distributions as a power-law distribution with the index $p$ which is assumed as a free parameter,
\begin{equation}
    \dot{N}_{e,\rm inj} = \dot{N}_0
    \begin{cases}
        \left( \frac{\gamma}{\gamma_{\rm min}}\right)^2 \ (\gamma < \gamma_{\rm min}) \\
        \left( \frac{\gamma}{\gamma_{\rm min}}\right)^{-p} \exp \left( -\frac{\gamma}{\gamma_{\rm max}}\right) \ (\gamma > \gamma_{\rm min} )
    \end{cases}, 
    \label{injection}
\end{equation}
where $\gamma_{\rm min}$ and $\gamma_{\rm max}$ are the minimum and maximum Lorentz factor, respectively. We estimate $n_e$ from Equation (\ref{energy_consv}) and calculate the normalization, $\dot{N}_0$, by $\dot{N}_e = n_e \pi R_j^2 c = \int^{\infty}_1 \dot{N}_{e, \rm inj} d\gamma$. We have assumed that the electron energy distribution for $\gamma < \gamma_{\rm min}$ is close to the thermal distribution. We set the maximum Lorentz factor as $\gamma_{\rm max} = {\rm min}( \gamma_{\rm cut}, \gamma_{\rm rad})$, where 
\begin{align}
    \gamma_{\rm cut} &= \left(\frac{\delta B}{B}\right)^2 \xi \sigma_\pm \nonumber \\
    &\simeq 1.0\times10^6 \ \left(\frac{\delta B/B}{0.33}\right)^2\xi_3\sigma_{\pm, 4},
    \label{max}
\end{align}
and $\gamma_{\rm rad}$ is the Lorentz factor estimated by balancing the acceleration and cooling timescales. We estimate the reconnection acceleration timescale as $t_{\rm acc} = \eta_{\rm acc}\gamma m_ec/(e B_{\rm dis})$,\footnote{If electrons are only accelerated by the reconnection electric field of the turbulent magnetic field, the acceleration timescale is estimated to be $t_{\rm acc} = \eta_{\rm acc}\gamma m_ec/(e \delta B)$. We estimate $\delta B$ as $\delta B \approx (\delta B/B)B_{\rm dis}$, and thus, the acceleration timescale is longer than that given in the text by a factor of $B/\delta B$. This affects the estimate of $\gamma_{\rm rad}$ but does not affect our results within the parameters of our interest.} where $\eta_{\rm acc} \sim 10$ is the numerical factor and $e$ is the electric charge \citep[e.g.,][]{Guo2020}. 
We estimate $B_{\rm dis}$ by Equation (\ref{after_dis}) as
\begin{align}
    B_{\rm dis} &= \sqrt{\frac{L_j}{1+0.5\sigma_{\rm dis}}\frac{4 \sigma_{\rm dis}}{R_j^2 c}} \nonumber \\ 
    &\simeq 1.7\ \dot{m}_{-4}^{1/2} M_9^{-1/2} R_{j,2}^{-1} \left(\frac{\sigma_{\rm dis}/(1+0.5\sigma_{\rm dis})}{0.66}\right)^{1/2} \rm G.
    \label{Bdis}
\end{align}

The minimum Lorentz factor is written by
\begin{align}
    \gamma_{\rm min} &= {\rm max} \left(1,\left( \frac{\delta B}{B}\right)^2 \frac{n_pm_p}{n_em_e}{\sigma_{\rm ent}}\right) \nonumber \\
    &\simeq 2.0\ \left(\frac{\delta B/B}{0.33}\right)^2\sigma_{{\rm ent},-2}. 
    \label{min}
\end{align}
We note that after plasma entrainment, the entrained plasma is the main component of the internal energy inside the jet. Thus, we need the factor $(n_p m_p)/(n_em_e)$ for estimating the magnetic energy density per electron \citep[cf.][]{Petropoulou+2019}.

We consider the cooling processes of the electrons as synchrotron radiation, synchrotron self-Compton scattering, and adiabatic cooling. As for the escape process, we consider that the electrons escape from the jet diffusely through resonant scattering with the turbulence induced by KHI. We explain how to estimate the cooling and the escape timescales in Appendix \ref{loss_jet}.

\section{Results for strong plasma entrainment} \label{sec:results}
\begin{figure}[tbp]
    \centering
    \gridline{\fig{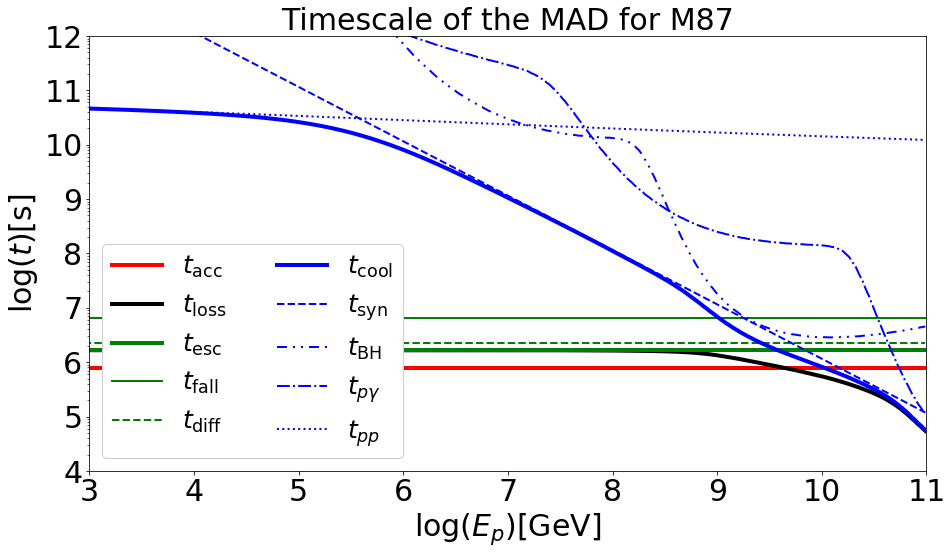}{1.0\linewidth}{}}
    \caption{Cooling (blue), escape (green), loss (black), and acceleration (red) timescales in the MAD as a function of the proton energy. $t_{\rm syn}$, $t_{\rm BH}$, $t_{pp}$, and $t_{p\gamma}$ are the synchrotron, Bethe-Heitler, $pp$ inelastic collision, and photomeson production cooling timescales, respectively. Due to the hard-sphere-like diffusion coefficient, the acceleration and the diffusion timescales (green dashed line) do not depend on the proton energy.}
    \label{time-MAD}
\end{figure}

In this section, we apply the Jet-MAD model to M87 and compare the multi-wavelength photon spectra obtained by the Jet-MAD model to the observational data. We tabulate the parameters and the physical quantities for M87 in Table \ref{MAD_parameter_M87}. We show the various timescales of the MAD model as a function of the proton energy $E_p$ for M87 in Figure \ref{time-MAD}. Due to the hard-sphere-like diffusion coefficient, the diffusion and acceleration timescales of the MAD model are independent of $E_p$. The cooling timescales are negligibly small for $E_p \lesssim 10^9 ~ \rm GeV$.

We estimate the index of the injection term for the MAD, $s_{\rm inj}$, by the stochastic acceleration process with steady-state and hard-sphere approximations. As long as the cooling timescale is negligible, this process results in $s_{\rm inj} = -(1/2) + \sqrt{(9/4) + \epsilon}$, where $\epsilon = t_{\rm acc}/t_{\rm diff}$ is the escape-acceleration timescale ratio \citep[see Section 2 of][]{Stawarz2008}. We obtain $s_{\rm inj} \simeq 1.1$ in our fiducial parameters.

\begin{table*}[t]
\centering
\caption{The list of the parameters and the physical quantities of the Jet-MAD model for M87.}\label{MAD_parameter_M87}
 Parameters of the Jet-MAD model for strong plasma entrainment
\begin{tabular}{lclcl}
\hline \hline 
Parameter & &&&  Description \\ \hline
$\dot{m}$ &&  $6.0\times 10^{-5}$ && Mass accretion rate normalized by $L_{\rm Edd}/c^2$\\
$M_{\rm BH} ~ [M_\odot]$ && $6.3\times 10^9$ && BH mass \\
$r ~ [R_g]$ && 10 && Size of the emission region for the MAD\\

$Q_i/Q_e$ && 10 && Ratio of the electron-proton heating rate \\
$\eta_{\rm turb}$ && 0.833 && Numerical factor of the effective mean free path \\
$\alpha$ && 0.3 && Viscous parameter \\
$\beta$ && 0.1 && Plasma beta \\
$\epsilon_{\rm NT}$ && 0.33 && Fraction of the nonthermal particle energy to the dissipation energy \\
$\epsilon_{\rm dis}$ && 0.15 && Fraction of the dissipation energy to the accretion energy \\ \hline
$\sigma_{\rm ent}$ && 0.01 && Magnetization parameter after plasma entrainment at the jet  dissipation region \\
$p$ && 2.16 && Index of the injection term for the jet \\
$\xi$ && $10^3$ && Numerical factor of the blob expansion \\
$\delta B/B$ && 0.35 && Ratio of the amplitude for the perturbed and the ordered magnetic fields inside the jet \\
$R_j ~ [R_g]$ && $10^2$ && Lateral size of the jet dissipation region \\
\hline \\
\end{tabular}

Derived physical quantities of M87
    \begin{tabular}{lclcl}
    \hline \hline  
    Physical quantity &&&& Description \\ \hline
       $s_{\rm inj}$ && 1.1 && Index of the injection term for the MAD \\
       $\sigma_\pm$ && $1.0\times 10^5$ && Magnetization parameter at the jet base \\
       $ B_{\rm dis} \rm [G]$ &&$6.5\times10^{-2}$ && Magnetic field strength at the jet dissipation region \\
       $n_e \rm \ [cm^{-3}]$ && 22 && Electron number density at the jet dissipation region\\
       $L_j \ \rm [erg~s^{-1}]$ && $2.8\times 10^{43}$ && Injected power at the jet base\\
       $L_e \ \rm [erg~s^{-1}]$ &&$1.7 \times 10^{41}$ && Electron luminosity at the jet dissipation region\\
       $\delta_D$ &&1.9 && Doppler factor\\
        \hline
    \end{tabular}
    \tablecomments{We use $\alpha$, $\beta$, $\epsilon_{\rm NT}$, and $\epsilon_{\rm dis}$ to estimate the physical quantities of the MAD, such as magnetic fields and the luminosities of the nonthermal particles (see Appendix \ref{MAD model}).}
\end{table*}

\begin{figure*}[tbp]
    \centering
    \gridline{\fig{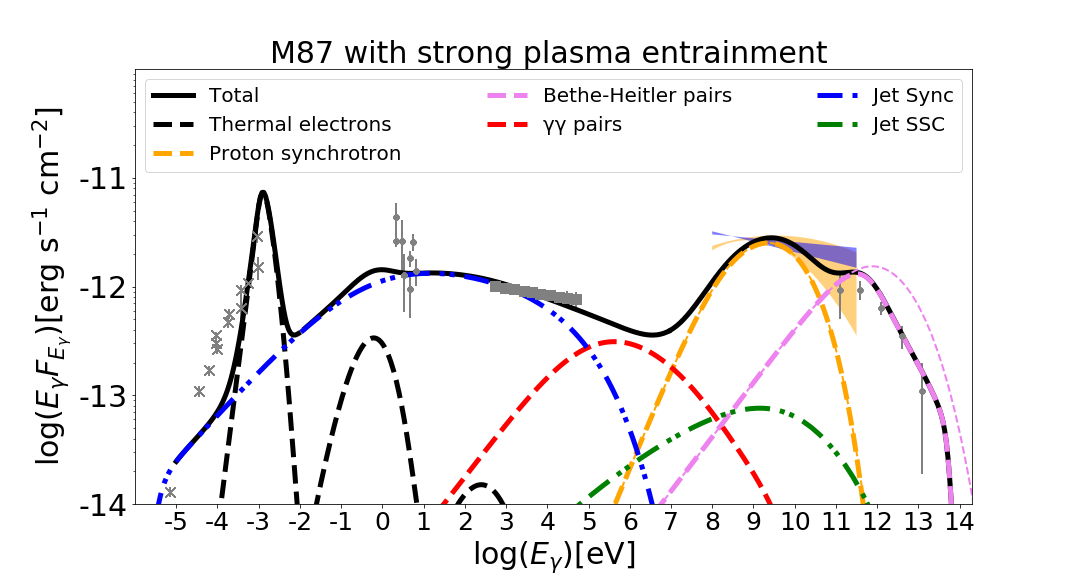}{\linewidth}{}}
    \caption{Photon spectrum for M87. The thick and thin lines are the photon spectra after and before internal attenuation by the Breit-Wheeler process, respectively. Gray points are the multi-wavelength simultaneous data taken from \citet{EHT_MWL2021} and \citet{MAGIC2020}. Blue and orange shaded regions are power-law and log-parabola fitting spectrum taken from the Fermi-4FGL-DR3 catalog (\url{https://fermi.gsfc.nasa.gov/ssc/data/access/lat/12yr_catalog/}).} 
    \label{spectrum_M87}
\end{figure*}

\begin{figure}
    \centering
    \gridline{\fig{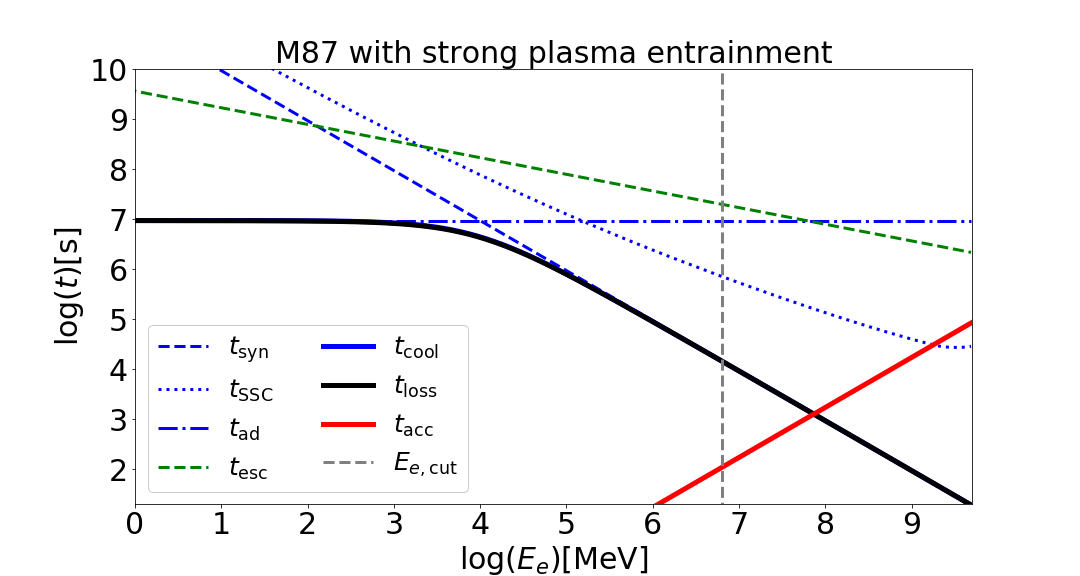}{1.1\linewidth}{}}
    \caption{Same as Figure \ref{time-MAD}, but timescales in the jet as a function of the electron energy.  $t_{\rm SSC}$ (dotted) and $t_{\rm ad}$ (dot-dashed) are the synchrotron self-Compton and adiabatic cooling timescales, respectively. The vertical dashed gray line represents $E_{e, \rm cut} = \gamma_{\rm cut}m_ec^2$.}
    \label{M87_timescale_jet}
\end{figure}

First, we consider a case with $(\delta B/B)^2 \sigma_{\rm ent} < 1$. We show the results in Figures \ref{spectrum_M87} and \ref{M87_timescale_jet}. Thermal synchrotron radiation reproduces the mm/sub-mm data. Synchrotron radiation by the nonthermal protons and secondary electron-positron pairs produced via the Bethe-Heitler process in the MAD reproduce the GeV and TeV gamma-ray data, respectively. Synchrotron radiation by the jet contributes to the optical, UV, and X-ray data. Our model can reproduce multi-wavelength data except for tens of GHz radio data. In our one-zone jet emission model, the optical depth of the synchrotron self-absorption at the GHz band is $\sim 1$. Higher-frequency radio data should be produced by more inner parts of the jet where the optical depth of the synchrotron self-absorption is higher. Thus, if we extend our one-zone model to a one-dimensional one to involve the emission from the inner parts of the jet, our model will be able to explain the radio data.

We estimate the photon energy emitted by electrons with $\gamma_{\rm max}$ to be
\begin{align}
E_{\gamma, \rm max} &= \gamma_{\rm max}^2\frac{heB_{\rm dis}}{2 \pi m_e c} \nonumber \\
&\simeq 2.4\times10^4 \dot{m}^{1/2}_{-4} M^{-1/2}_{9}R_{j,2}^{-1}\left(\frac{\delta B/B}{0.33}\right)^4 \nonumber \\
&\times \xi^2_{3} \sigma^2_{\pm, 4} \left(\frac{\sigma_{\rm dis}/(1+0.5\sigma_{\rm dis})}{0.66}\right)^{1/2} \ \rm eV,
\label{Egam_max}
\end{align}
 where $h$ is the Planck constant. Thus, the electrons accelerated by the magnetic reconnection have a Lorentz factor high enough to emit the optical to the X-ray photons via synchrotron radiation. On the other hand, the photon flux of the synchrotron self-Compton scattering is much lower than the synchrotron flux as shown in Figure \ref{spectrum_M87}. In our model, synchrotron flux is calibrated by the optical data to be $F_{\rm opt} \simeq 8\times 10^{-13} \ \rm erg~s^{-1}~cm^{-2}$, and we estimate the photon energy density as $U_{\rm rad} = F_{\rm opt} D^2/(R_j^2 c) \simeq 8.4\times 10^{-6} \ \rm erg ~ cm^{-3}$, where $D = 17 ~ \rm Mpc$ is the distance to M87. We also estimate the magnetic energy density to be $U_B = B_{\rm dis}^2/(8\pi) \simeq 1.7 \times 10^{-4}\ \rm erg~cm^{-3}$. The energy density ratio is estimated to be $U_{\rm rad}/U_B \simeq 0.05$, and thus the photon flux of the synchrotron self-Compton scattering is much lower than the synchrotron component. 

Figure \ref{M87_timescale_jet} shows the various timescales in the jet as a function of the electron energy. For our fiducial parameters, electrons are efficiently scattered by the turbulence, and the diffusive escape is inefficient. The nonthermal electrons are mainly cooled by the adiabatic expansion and the synchrotron radiation for $\gamma < 1.9\times 10^4$ and $\gamma > 1.9\times 10^4$, respectively. The photon energy emitted by the electrons with $t_{\rm ad} = t_{\rm syn}$ is estimated to be $E_{\gamma,\rm cool} \simeq 0.29 ~ \rm eV$, at which the synchrotron spectrum is the peak, and that enables our model to explain the optical data.

Next, we consider the case with $(\delta B/B)^2 \sigma_{\rm ent} > 1$. In this case, $\gamma_{\rm min}$ is always higher than that of $t_{\rm ad} = t_{\rm syn}$, and then, the system is in the fast cooling regime. Also, we estimate the electron energy density as $U_e \approx \gamma_{\rm min}n_em_ec^2 = \left(\delta B/B\right)^2 \sigma_{\rm ent} n_pm_p c^2 = U_p$. Then, the electron luminosity is the same as the proton luminosity, and we estimate the electron luminosity to be $L_e = (L_j/2)/(1+0.5\sigma_{\rm dis})$ by Equation (\ref{after_dis}). The electron luminosity is restricted to be $L_e \simeq L_j\times10^{-3} $ by the optical data, and we give $\sigma_{\rm dis} \simeq 10^3$. Since $(\delta B/B)^2 \sigma_{\rm ent} > 1$ and $\sigma_{\rm dis} \simeq 10^3$, we obtain $(\delta B/B) \simeq 10^{-1.5}$ by Equation (\ref{sig_dis}). Then, we estimate the maximum photon energy to be $E_{\gamma,\rm max} \simeq 1.1\times10^2~ \rm eV$, which is low to emit the X-rays. Thus, we conclude that $\left(\delta B/B\right)^2 \sigma_{\rm ent} > 1$ is unlikely to explain the multi-wavelength data.

\section{Resutlts for weak or no plasma entrainment}\label{entrainemt}
So far, we assumed that the KHI occurs at the edge of the jet, which induces the plasma entrainment from the ambient gas. In this section, we investigate the cases of weak or no plasma entrainment from the ambient gas. 

First, we discuss the case with weak plasma entrainment. We define `weak plasma entrainment' such that the entrained plasma is the main component of the internal energy density inside the jet ($(n_p m_p)/(n_e m_e)>1$), whereas the entrained plasma is not the main component of the electron number density ($n_e > n_p$). 

\begin{figure*}[tbp]
    \centering
    \gridline{\fig{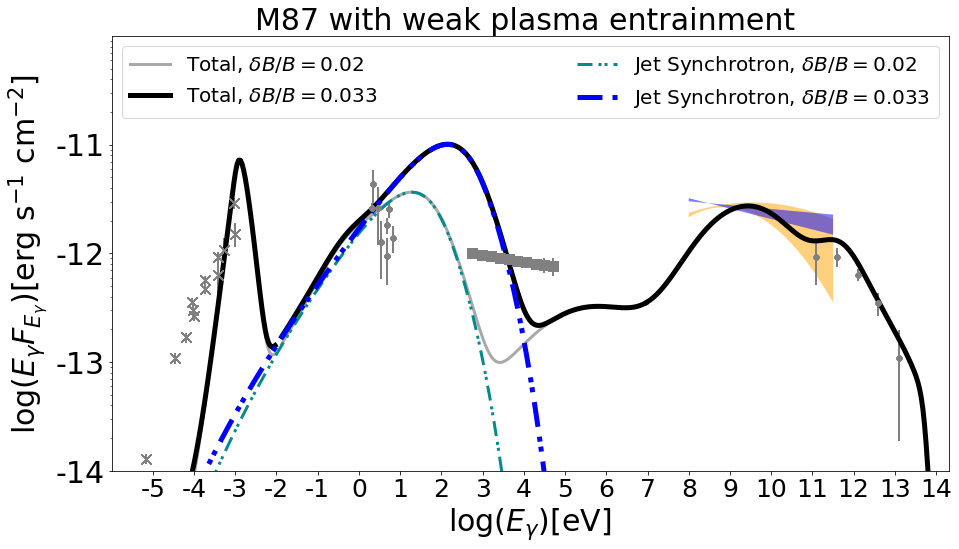}{1.0\linewidth}{}}
    \caption{Same as Figure \ref{spectrum_M87}, but for the weak plasma entrainment. Total and jet synchrotron photon spectra with $\delta B/B = 0.033$ (thick lines) and $\delta B/B = 0.02$ (thin lines). We set $\delta B/B = 0.02$ to emit the optical photons and $\delta B/B = 0.033$ to emit the X-ray photons. Both photon spectra cannot reproduce the multi-wavelength data simultaneously. Due to high $\sigma_{\rm dis}$, the photon flux of the synchrotron self-Compton scattering is low.}
    \label{wo-plasma}
\end{figure*}

As $(n_p m_p)/(n_e m_e)>1$, we estimate the proton number density by Equation (\ref{energy_consv}), and then, $\sigma_{\rm ent} > (m_e/m_p)\xi \sigma_\pm \simeq 10^5$ is required to satisfy $n_e > n_p$. We estimate the photon energy emitted by electrons with $\gamma_{\rm min}$ to be 
\begin{align}
    E_{\gamma,\rm min} &= \gamma_{\rm min}^2\frac{heB_{\rm dis}}{2 \pi m_e c} \nonumber \\
    &\simeq 8.0\times10^6 ~\dot{m}^{1/2}_{-4} M^{-1/2}_{9} R_{j,2}^{-1}\left(\frac{\delta B/B}{0.33}\right)^4 \nonumber \\
&\times \sigma_{\rm ent,5}^2 \left(\frac{\sigma_{\rm dis}/(1+0.5\sigma_{\rm dis})}{0.66}\right)^{1/2}~\rm eV.
\label{Egammin}
\end{align} 
The obtained minimum Lorentz factor is too high to emit the optical and X-ray photons. 
Based on Equation (\ref{Egammin}), $\delta B/B \sim 0.01$ is required to explain the X-ray data. Even if we tune $\delta B/B$ to emit the X-ray photons by electrons with $\gamma_{\rm min}$, the photon spectrum is too peaky to fit the optical and the X-ray data simultaneously. 
 For the case of weak plasma entrainment, we show the result in Figure \ref{wo-plasma} and tabulate the changed parameters and physical quantities in Table \ref{dif_parameter}. Other parameters and physical quantities are the same as those tabulated in Table \ref{MAD_parameter_M87}. We conclude that the weak plasma entrainment is disfavored to explain the multi-wavelength data.
\begin{table*}[t]
\centering
\caption{The list of the changed parameters and the physical quantities for the case of the weak plasma entrainment.}\label{dif_parameter}
\begin{tabular}{cccc|ccc}
\hline
 & $\sigma_{\rm ent}$ & $\sigma_{\rm dis}$ & $\delta B/B$& $B_{\rm dis} ~ [\rm {G}]$ & $n_e~ \rm [cm^{-3}]$ & $L_e ~\rm [erg ~ s^{-1}]$ \\ \hline
Thick &$5.5\times10^4$& $9.2\times10^2$& $0.033$ &$0.92$ & $8.2\times10^{-4}$ & $6.4\times10^{40}$\\ 
Thin &$5.5\times10^4$& $2.5\times10^3$&$0.02$ & $0.92$ & $8.2\times10^{-4}$&$2.4\times10^{40}$\\ 
\hline 
\end{tabular}
\end{table*}

Next, we discuss the case with no plasma entrainment. In this case, the magnetic reconnection accelerates all the electrons up to $\gamma \sim (\delta B/B)^2\xi \sigma_\pm \simeq 10^7$. This value is also too high to emit the optical and X-rays. Thus, we conclude that no plasma entrainment is disfavored to explain the multi-wavelength data for the same reason as the weak plasma entrainment.

\section{Disucussion}\label{sec:discuss} 
\subsection{Particle energy distribution: power-law tail}\label{sub:tail}
Some recent PIC simulations of high $\sigma$ magnetic reconnection exhibit a power-law tail at $\gamma > \sigma$ in the nonthermal particle spectrum \citep{Hakobyan2021, Hakobyan2023}. In this subsection, we take the power-law tail into account on our model and rewrite the injection term as
\begin{equation}
      \dot{N}_{e,\rm inj} \propto
    \begin{cases}
        \left( \frac{\gamma}{\gamma_{\rm min}}\right)^2 \ (\gamma < \gamma_{\rm min}) \\
        \left( \frac{\gamma}{\gamma_{\rm min}}\right)^{-p} \ (\gamma_{\rm min}< \gamma < \gamma_{\rm cut}  ) \\
        \left( \frac{\gamma}{\gamma_{\rm cut}}\right)^{-p_{\rm tail}} \exp \left(-\frac{\gamma}{\gamma_{\rm rad}}\right) \ (\gamma > \gamma_{\rm cut})\\
    \end{cases}, 
\label{tail}
\end{equation}
where $p_{\rm tail}$ is the spectral index for the electrons higher than $\gamma_{\rm cut}$. We use $p_{\rm tail} = $ 2.5 and 3.0. We evaluate the effects on the multi-wavelength photon spectrum from the jets in Section \ref{tail-jet} and on the plasma loading in Section \ref{tail-base}.

\subsubsection{Power-law tail in the jet dissipation}
\label{tail-jet}
We apply the Jet-MAD model with the power-law tail distribution to M87 and find that the Jet-MAD model can explain the multi-wavelength data even if the differential electron number spectrum has the power-law tail. The index of synchrotron flux is $s \approx (2-p)/2$ for the fast cooling regime. We obtain $s = -0.25$ and $-0.5$ for $p= 2.5$ and 3.0, respectively, with which the synchrotron flux emitted by the electrons in the tail ($\gamma> \gamma_{\rm cut}$) has little effect on the overall photon spectrum. Even if the tail index is a bit hard, the photon energy higher than $E_{\gamma, \rm max}$ does not affect the photon spectrum (see Figure \ref{spectrum_M87}). We conclude that our model can explain the multi-wavelength data independent of the uncertainty for the reconnection acceleration.

\subsubsection{Power-law tail for the electron-positron pair injection into the jet} \label{tail-base}
\begin{figure}
    \centering
    \gridline{\fig{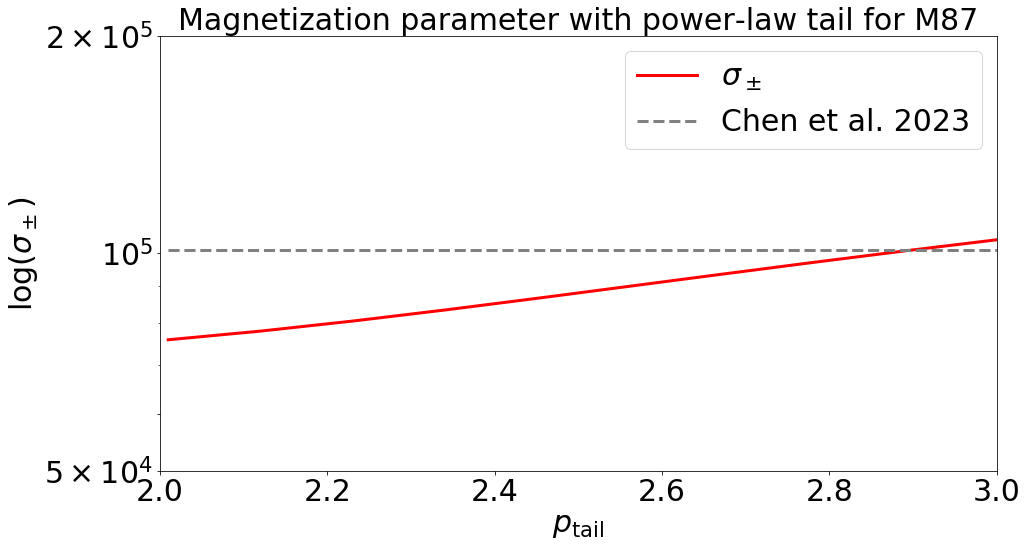}{1.0\linewidth}{}}
    \caption{Estimated magnetization parameter for the power-law tail distribution as a function of the tail index (red solid line) and the magnetization parameter estimated in the way of \citet{Chen+2022} (gray dashed line).}
    \label{fig:enter-label}
\end{figure}

Next, we evaluate the effect of the power-law tail on the electron-positron pair injection at the jet base. With the power-law tail, the photon spectrum extends up to a higher energy than that without the tail. One might expect a high pair production efficiency, compared to that estimated with the exponential cutoff as in \citet{Chen+2022}. We estimate the pair production rate and magnetization parameter at the jet base with the power-law tail distribution (see Appendix \ref{tail_para} for details). We find that the estimated magnetization parameters with the tail are comparable to that in our fiducial treatment: $\sigma_\pm = 8.8\times 10^4$ and $1.0 \times 10^5$ for $p_{\rm tail} = 2.5$ and $3.0$, respectively. We show the estimated magnetization parameter as a function of the tail index in Figure \ref{fig:enter-label}. As shown in Figure \ref{fig:enter-label}, the estimated magnetization parameter hardly changes its value with the tail index. This comes from the self-regulation of the synchrotron spectrum to peak at $E_{\gamma, \rm cut, ms} \sim m_e c^2$. Therefore, the power-law tail does not affect our conclusions.

\subsection{Acceleration by the Alfv\'en wave dissipation} \label{sec:Alf}
In this subsection, we discuss the Alfv\'en waves dissipation as an alternative particle acceleration mechanism in the jet \citep{Nattila2022}. Alfv\'en waves are produced by the magnetic reconnection in the BH magnetosphere or by the interaction of the jet with disk wind. These Alfv\'en waves propagate along the magnetic field lines and split into forward sound waves and backward Alfv\'en waves by the parametric decay instability (PDI) \citep{GaleevOraevskii1963, SagdeevGaleev1969, Derby1978}. We consider that the Alfv\'en wave turbulence is generated by the interaction of the forward Alfv\'en waves with backward Alfv\'en waves at the dissipation region, and the dissipation of Alfv\'en wave turbulence accelerates the electrons. It is shown that in a highly magnetized medium, the PDI can split the Alfv\'en waves into backward Alfv\'en waves and forward sound waves only when $(\delta B/B)^2 \sigma \leq 1 $ is satisfied \citep{Ishizaki2024} \footnote{If the Alfv\'en waves are in $(\delta B/B)^2 \sigma \geq 1$, they can split into backward Alfv\'en and backward sound waves (Ishizaki, private communication). In this case, due to the high growth rate of the PDI, the dissipation region is much smaller than that we consider in this paper, resulting in the high optical depth of the synchrotron self-absorption. If we tune $\sigma_{\rm ent}$ to avoid the high optical depth of the synchrotron self-absorption, our model with the Alfv\'en wave dissipation scenario may explain the multi-wavelength data.}. We set $\delta B/B = 1/\sqrt{\sigma_\pm} \simeq 3\times 10^{-3}$ which is close to the maximum value of $\delta B/B$ that satisfies the PDI growth condition. Then, the maximum Lorentz factor is written as $\gamma_{\rm max} \approx \xi$ by Equation (\ref{max}), and the value is lower than that of the magnetic reconnection scenario.

\begin{figure}[t]
    \centering
    \gridline{\fig{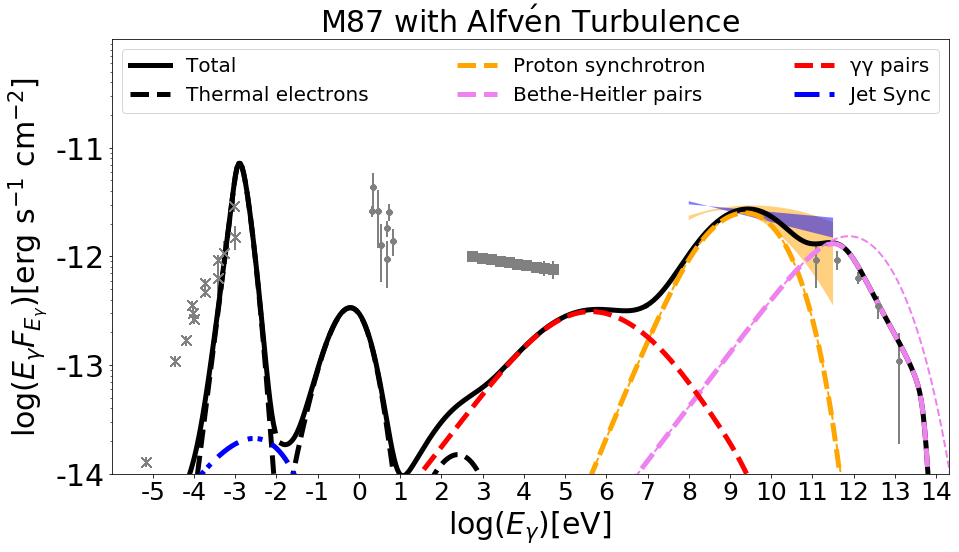}{1.0\linewidth}{}}
    \caption{Same as Figure \ref{spectrum_M87}, but for the acceleration by the Alfv\'en wave dissipation. We set the power-law index as 1.3 to achieve the high synchrotron photon flux.}
    \label{spectrum_M87_Alfven}
\end{figure}

\begin{figure}
    \centering
    \gridline{\fig{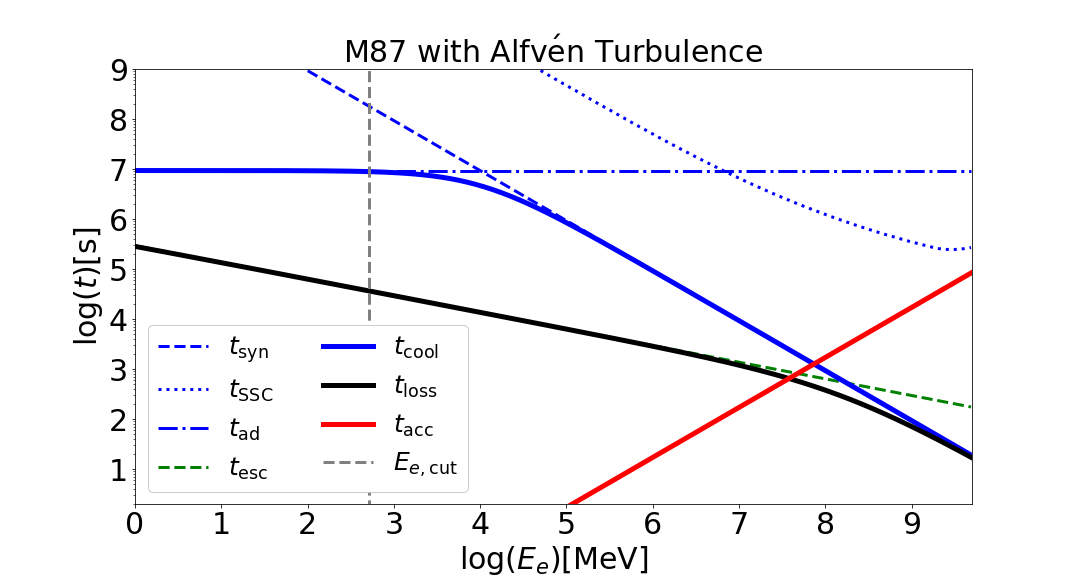}{1.1\linewidth}{}}
    \caption{Same as Figure \ref{M87_timescale_jet}, but for the acceleration by the Alfv\'en wave dissipation.}
    \label{M87_timescale_AW}
\end{figure}

We show the results in Figures \ref{spectrum_M87_Alfven} and \ref{M87_timescale_AW}. We estimate the synchrotron photon energy emitted by electrons with $\gamma_{\rm max}$ to be
\begin{align}
  E_{\gamma, \rm max} &= \gamma_{\rm max}^2 \frac{heB_{\rm dis}}{2\pi m_ec}\nonumber \\ 
  &\approx  2\times10^{-2} \dot{m}_{-4}^{1/2} M_9^{-1/2} \xi_3^2 R_{j,2}^{-1} \nonumber \\
  &\times \left(\frac{\sigma_{\rm dis}/(1+0.5\sigma_{\rm dis})}{0.66}\right)^{1/2}  \rm eV,
  \label{max_Alf}
\end{align}
which is too low to emit X-rays. $M_{\rm BH}$ and $\dot{m}$ are fixed by the observational data, and $\sigma_{\rm dis}/(1+0.5\sigma_{\rm dis}) \leq 2$ is generally satisfied. To emit the X-rays, $\xi > 10^5$ or $R_j < R_g$ are required, but both are highly unlikely to be satisfied.

Diffusive escape also affects the photon spectrum. For the Alfv\'en wave dissipation scenario, the amplitude of the turbulent magnetic field is lower than that of the magnetic reconnection scenario. Due to the low amplitude, the pitch angle scattering rate is low, leading to a high spatial diffusion coefficient. The diffusive escape timescale is always shorter than the adiabatic and synchrotron cooling timescales, causing a low radiation efficiency, compared to the magnetic reconnection scenario (see Figure \ref{M87_timescale_AW}). As a result, the calculated photon flux is low for the Alfv\'en wave dissipation scenario. We conclude that the dissipation of the Alfv\'en wave turbulence is disfavored as the acceleration mechanism inside the jet.

\subsection{Neutrino luminosity of the Jet-MAD model}
Here, we roughly estimate neutrino emission from M87 based on our Jet-MAD model. Protons are entrained into the jet from the ambient gas via the KHI. The magnetic reconnection accelerates the entrained protons inside the jet, while the MHD turbulence also accelerates the protons inside the MAD. The accelerated protons can produce neutrinos via $pp$ inelastic collisions ($p+p \rightarrow p+p+\pi$) or photomeson production ($p+\gamma \rightarrow p+ \pi$). We discuss neutrinos from the jet in Section \ref{Neu-Jet} and the MAD in Section \ref{Neu-MAD}. 

\subsubsection{Neutrino luminosity from Jet}\label{Neu-Jet}
In this subsection, we estimate the neutrino emission from jets. The number density of the protons inside the jet is about $22 \ \rm cm^{-3}$ (see Table \ref{MAD_parameter_M87}). We estimate the $pp$ inelastic collision timescale as $t_{pp} = 1/(n_p \sigma_{pp} \kappa_{pp} c) \simeq 5.0 \times 10^{13} \ \rm s$, where $\sigma_{pp} \approx 60~ \rm mb$ is the cross section and $\kappa_{pp} \approx 0.5$ is the inelasticity of the $pp$ inelastic collisions, respectively. The $pp$ inelastic collision timescale is much longer than the adiabatic cooling timescale $t_{\rm ad} = 3R_j/c \simeq 9.4 \times 10^6 \ \rm s$, and thus, neutrino production via $pp$ inelastic collisions is inefficient due to the low proton number density. $t_{pp}$ is also much longer than photomeson production timescale, $t_{p\gamma}$, as shown below. Thus, we ignore the effect of $pp$ inelastic collisions here.

We estimate neutrino flux produced by the photomeson production. First, we estimate the maximum Lorentz factor of the protons. At the initial stage of the plasma entrainment, protons are a subdominant component of the internal energy (i.e., $n_em_e \gtrsim n_p m_p$). Then, the proton magnetization parameter is $\sigma_p \simeq\sigma_\pm $, and the magnetic reconnection can achieve $\gamma_p \sim (\delta B/B)^2\xi \sigma_\pm  \simeq 10^7$, where $\gamma_p$ is the proton Lorentz factor. Neutrino energy is estimated to be $E_{\nu} \approx 0.05 E_p $. Then, $E_{\nu} \simeq 5.0\times10^{14} ~ \rm eV$ for $E_p \simeq 1.0\times10^{16} ~ \rm eV$. For the photomeson production, the protons interact with the photons of $ E_\gamma \simeq 0.3 ~\rm GeV$ in the proton rest frame for the delta resonance approximation. Then, the protons of $\gamma_p \simeq 10^7$ interact with $E_{\gamma} \simeq 30 ~\rm eV$, which is in the UV band. We estimate the number density of the UV photons inside the jet as $n_{\rm UV} \approx F_{\rm UV} D^2 / (R_j^2 c \epsilon_{\rm UV}) \simeq 2.8\times 10^5 \ \rm cm^{-3}$, where $F_{\rm UV} = 1.3\times10^{-12}\ ~ \rm erg~s^{-1}~cm^{-2}$ is the flux of the UV photons, and $\epsilon_{\rm UV} \simeq 30 \ \rm eV$ is the UV photon energy. We estimate the photomeson production timescale as $t_{p \gamma} = 1/(n_{\rm UV} \sigma_{p\gamma}\kappa_{p\gamma}c) \simeq 1.2\times10^{12} ~\rm s$, where $\sigma_{p\gamma}\approx 0.5 ~\rm mb$ is the cross section and $\kappa_{p\gamma} \approx 0.2$ is the inelasticity of the photomeson production, respectively. By equation (\ref{after_dis}), we estimate the proton luminosity as $L_p = L_j/(1+0.5\sigma_{\rm dis}) - L_e \simeq L_j$. Then, we optimistically estimate the luminosity of the neutrino as $L_\nu \approx f_{p\gamma} L_p \simeq  5.8\times 10^{37} ~\rm erg ~ s^{-1}$, where $f_{p\gamma} = t_{\rm loss}/t_{p\gamma} \simeq 1.2\times 10^{-7}$ is the pion production efficiency. Regarding the loss timescale, diffusive escape is dominant, compared to the adiabatic cooling, in the relevant energy (see Figure \ref{M87_timescale_jet}), and the proton synchrotron cooling timescale is negligibly long. The neutrino flux is estimated to be $F_\nu = L_\nu/(4\pi D^2) \simeq 1.7\times 10^{-15} \ \rm erg~s^{-1}~cm^{-2}$, which is challenging to detect with near-future neutrino detectors.

\subsubsection{Neutrino luminosity from MAD}\label{Neu-MAD}
Next, we estimate the neutrino emission from MADs. As shown in Figure \ref{time-MAD}, the proton maximum Lorentz factor is estimated to be $\gamma_{p,\rm max} \simeq 4.8\times 10^{9}$. We estimate the proton number density to be $n_p \approx \rho/m_p \simeq 4.1 \times 10^4 \ \rm cm^{-3}$, where $\rho$ is the mass density, and then, the $pp$ inelastic collision timescale is estimated to be $t_{pp} \simeq 2.7 \times 10^{10} \ \rm s$. The $pp$ inelastic collision timescale is longer than the diffusion timescale $t_{\rm diff} \simeq  2.2\times10^6 \ \rm s$. $t_{pp}$ is also much longer than photomeson production timescale, $t_{p\gamma}$, as shown below. Thus, protons emit neutrinos via photomeson production. 

We estimate the neutrino flux of the photomeson production in the same way as the emission from the jet. The protons of $\gamma_{p, \rm max} \simeq 4.8 \times 10^{9}$ interact with $6.3\times10^{-2}$ eV photons, and we estimate the number density of the infrared photons as $n_{\rm IR} \approx F_{\rm IR} D^2 / (R_d^2 c \epsilon_{\rm IR}) \simeq 4.9\times 10^8 \ \rm cm^{-3}$, where $F_{\rm IR} = 4.7\times10^{-14} ~ \rm erg~s^{-1}~cm^{-2}$ is the flux of the infrared photons emitted by the Comptonization process and $\epsilon_{\rm IR} \simeq 6.3\times 10^{-2} \ \rm eV$ is the infrared photon energy. Then, the photomeson production timescale is estimated to be $t_{p \gamma} \simeq 6.9\times10^{8}~ \rm s$. The nonthermal proton luminosity of the MAD is estimated to be $L_p(\gamma_{p,\rm max}) \approx \epsilon_{\rm NT} \epsilon_{\rm dis} \dot{M}c^2 \exp(-1)\simeq 8.7\times 10^{41} \ \rm erg ~ s^{-1}$ (see Table \ref{MAD_parameter_M87}). We estimate the luminosity of the neutrino and the pion production efficiency as $L_\nu \approx f_{p\gamma} L_p(\gamma_{p,\rm max}) \simeq  1.3\times 10^{39} ~\rm erg ~ s^{-1}$ and $f_{p\gamma}  \simeq 1.5\times 10^{-3}$, respectively. We estimate the neutrino flux from the MAD to be $F_\nu \simeq 3.7\times 10^{-14} \ \rm erg~s^{-1}~cm^{-2}$. We also numerically compute the neutrino flux using the method given in Section 3 of \citet{Kimura2020}, which is consistent with our analytic estimate within a factor of 1.5. Protons of $\gamma_{p,\rm max} \simeq 4.8\times 10^{9}$ effectively produce the neutrinos of $2.4\times 10^{17} ~ \rm eV$. To observe the sub-EeV neutrinos, experiments that detect radio signals from the neutrino-induced particle shower, such as GRAND \citep{GRANDO2020}, RNO \citep{Aguilar+2021}, and Gen2-Radio \citep{Gen2-Radio2021}, are important.

Nevertheless, we conclude that it is still challenging to observe the high-energy neutrinos by the near-future detectors because the estimated neutrino flux is about 5 times lower than the sensitivity of the IceCube-Gen2 \citep{IceCubeGen22021}.

\subsection{Effect of the pitch angle anisotropy}
\citet{ComissoSironi2019} and \citet{Comisso+2020} showed that the electrons accelerated in the strongly magnetized turbulent plasma have an energy-dependent anisotropic particle distribution. For low-energy electrons, the reconnection electric field parallel to the guide field accelerates the electrons, and thus, their pitch angle is small. High-energy electrons have isotropic pitch angle distribution since the electrons are isotropized and stochastically accelerated by the pitch angle scattering. Thus, the pitch angle depends on the electron energy based on their simulations. Such an energy-dependent pitch angle distribution affects the multi-wavelength synchrotron spectrum \citep{Sobacchi+2021, GotoAsano2022}. In this paper, we ignore this effect. For the Jet-MAD model, we assume that the magnetic reconnection in a turbulent medium accelerates the electrons at a certain magnetization parameter. Thus, the electrons might have a small pitch angle distribution at the initial stage of the dissipation process. However, electrons can be scattered by the turbulence generated by the KHI, and then, the electrons could be isotropized when they emit optical and X-ray photons. Modeling with the pitch angle distribution is left as a future work.

\subsection{Time variability analysis}\label{sec_tb}
Time variability helps us to test our model. Our model predicts that optical and X-rays are emitted from the jet, while mm/sub-mm and gamma rays are emitted from the MAD. Based on the size of each emission region, we roughly estimate the variability timescales to be $10^6\;$s for the jet and $10^5\;$s for the MAD. Thus, pointing multi-wavelength instruments to M87 for more than one day is needed to test our model. Then, high-time resolution observations at least higher than $10^5\;$s such as mm/sub-mm observation with ALMA, infrared and optical observations with JWST and HST, X-ray observations with CHANDRA or Swift XRT, and TeV gamma-ray observations with MAGIC, HESS, and CTA \citep{CTA2011}, are desired. However, the sensitivity of the Fermi-LAT, which observes the GeV gamma rays, is insufficient to detect the time variability \citep{Atwood2009}. Our model predicts that mm/sub-mm and TeV gamma-ray light curves strongly correlate with day-scale time variability. Also, optical and X-ray light curves should correlate strongly with week-scale variability. We should note that the time variability cannot distinguish between strong and weak plasma entrainments since we consider the same dissipation region inside the jet. Future multi-wavelength observation campaigns will provide a test of our model.

\section{Conclusion} \label{sec:summary}
We construct the two-zone multi-wavelength emission model, the `Jet-MAD model', that takes into account the plasma loading to the jet based on the model proposed by \citet{Kimura+2022} and \citet{Chen+2022}. In the MAD, the MHD turbulence heats up the thermal electrons that emit thermal synchrotron radiation. The MHD turbulence also accelerates the nonthermal protons, which emit multi-wavelength photons via synchrotron radiation. For the jet, we consider that the velocity difference between the jet and the ambient gas generates turbulence via the KHI. This turbulence induces the entrainment of the ambient gas into the jet. The turbulence also triggers the magnetic reconnection inside the jet, accelerating the electrons that efficiently emit multi-wavelength photons. In this scenario, the Jet-MAD model can explain the simultaneous multi-wavelength data for M87 with a parameter set given in Table \ref{MAD_parameter_M87}. Synchrotron radiation from the jet explains the optical, UV, and X-ray data, and synchrotron radiation of the thermal electrons and protons from the MAD explains the mm/sub-mm and GeV gamma-ray data, respectively.

We examine the case of $(\delta B/B)^2\sigma_{\rm ent} > 1$. In this case, we obtain $\delta B/B \simeq 10^{-1.5}$ due to the electron luminosity restriction by the optical data. Then, the photon energy emitted by the electrons with $\gamma_{\rm max}$ is too low to emit the X-ray photons, and our model cannot explain the multi-wavelength data. Thus, we conclude that strong plasma entrainment satisfying the $(\delta B/B)^2 \sigma_{\rm ent} < 1$ is necessary to explain the multi-wavelength data.

We investigate the cases of weak or no plasma entrainment from the ambient gas. With weak or no plasma entrainment, magnetic reconnection accelerates the electrons up to $\sim 10^7$ with our fiducial parameter set. These electrons efficiently emit gamma rays but cannot explain the optical and X-ray data. Even if we tune the parameters such that the accelerated electrons emit the optical and X-ray photons, the photon spectrum is too peaky to reproduce the optical and X-ray data simultaneously. Thus, we conclude that the cases with weak or no plasma entrainment are disfavored to explain the multi-wavelength data. 

We examine the dissipation of the Alfv\'en wave turbulence as an alternative nonthermal electron acceleration mechanism. In this acceleration mechanism, we cannot explain the multi-wavelength data because the maximum energy of the electrons is too low to reproduce the observed X-ray data due to the low amplitude of the turbulent magnetic field. We conclude that the Alfv\'en wave dissipation scenario is disfavored as the acceleration mechanism.

These results suggest that the jet consists of the electron-positron pairs produced by the Breit-Wheeler process at the jet base and the electron-proton plasma entrained from the ambient gas via the KHI. These results also suggest that the plasma entrainment is strong enough to satisfy $(\delta B/B)^2 \sigma_{\rm ent} < 1$ and the magnetic reconnection scenario likely works as the nonthermal particle acceleration mechanism inside the jet. Our conclusion remains the same even if we have a power-law tail component for $\gamma > \gamma_{\rm cut}$. Time variability analysis should test our model. Our model predicts a strong correlation between optical and X-ray variabilities, and the strong correlation should also appear in mm/sub-mm and gamma-ray variabilities.

The optical and X-ray data require a soft power-law index of the electron energy distribution. Based on our assumption that the injection spectrum is the superposition of the accelerated electron distributions by the magnetic reconnection with various magnetization parameters, the soft power-law index implies that the magnetic energy dissipates more efficiently in a lower magnetization parameter environment. This is in line with our KHI-induced energy dissipation scenario. We consider that the KHI induces the entrainment of the ambient gas first. Then, the magnetization parameter decreases, and the plasma would be able to perturb the magnetic fields more, which induces the magnetic reconnection. Then, $\sigma$ becomes lower, and the plasma can perturb the magnetic field more, causing more energy dissipation in the low $\sigma$ environment.

 In this paper, we focus on the case study of M87. We can test our model by applying it to other radio galaxies, such as NGC 1275. This kind of test will make the applicability of our model more robust, but the detailed analysis and discussion will be left as future work.

\section*{acknowledgments}
We would like to thank Wataru Ishizaki and Hayk Hakobyan for their useful comments. The codes used for this study are available from R.K. or S.S.K. upon reasonable request. Discussions for this paper were performed many times at Science Lounge of FRIS CoRE. This work is partly supported by JSPS KAKENHI grant Nos. 22K14028, 21H04487, and 23H04899 (S. S. K.). This work is also supported by JST SPRING, grant No. JPMJFS2114 (R.K.). S.S.K. acknowledges support by the Tohoku Initiative for Fostering Global Researchers for Interdisciplinary Sciences (TI-FRIS) of MEXTs Strategic Professional Development Program for Young Researchers.

\appendix
\section{Detailed calculation method in the MAD model}\label{MAD model}
 In this appendix, we explain the detailed calculation method in the MAD model. First, we explain how to determine the electron temperature in the MAD model. As explained in  Section \ref{subsec:MAD}, we determine the temperature of the thermal electrons by equating the heating rate with the cooling rate:
\begin{equation}
    L_{\rm thermal} \approx \frac{Q_i}{Q_e}(1-\epsilon_{\rm NT})\epsilon_{\rm dis} \dot{M}c^2,
\end{equation}
where $L_{\rm thermal}$ is the luminosity of the photons emitted by the thermal electrons. As the electron heating rate, we use the description of \citet{Kawazura2020} which shows that when the compressive wave is dominant, the ratio of the electron-proton heating rate is approximated to be compressive-to-Alfv\'enic wave power ratio:
\begin{equation}
\frac{Q_i}{Q_e} \approx \frac{P_{\rm compr}}{P_{\rm AW}},
\end{equation} 
where $P_{\rm compr}$ is the compressive wave power and $P_{\rm AW}$ is the Alfv\'enic wave power. The wave power ratio in the MAD has not been established yet. We give $Q_i/Q_e$ as a parameter and set it as 10. Since the accretion flow is in the MAD state, the plasma beta is likely lower than 1. Hence, $Q_i/Q_e \simeq 10$ corresponds to $R_{\rm low} \simeq 10 $ in the $R_{\rm low}- R_{\rm high}$ prescription \citep[see][]{Moscibrodzka2016}, and $R_{\rm low}\simeq10$ is favored by the polarization observation of M87 \citep{EHT2021VIII}.

Next, we explain how to calculate the energy distributions of the nonthermal particles. To obtain the particle energy distributions, we solve the energy transport equation with the one-zone and steady-state approximations:
\begin{equation}
    -\frac{d}{dE_i} \left(\frac{E_i N_{E_i}}{t_{i,\rm cool}} \right) = \dot{N}_{E_i, \rm inj} - \frac{N_{E_i}}{t_{\rm esc}},
    \label{transport_MAD}
\end{equation}
where $N_{E_i}$ is the differential number spectrum, $t_{i, \rm cool}$ is the cooling time, $t_{\rm esc}$ is the escape time, $\dot{N}_{E_i, \rm inj}$ is the injection terms, and $i$ is the particle species. We write the injection terms as
\begin{equation}
    \dot{N}_{E_i, \rm inj} \approx \dot{N}_{0,\rm mad} \left( \frac{E_i}{E_{i, \rm cut}}\right)^{-s_{\rm inj}}\exp \left(-\frac{E_i}{E_{i, \rm cut}} \right).
    \label{injection_MAD}
\end{equation}
We estimate the normalization, $\dot{N}_{0,\rm mad}$, as $L_{p,\rm mad} = \int \dot{N}_{E_p, \rm inj} E_pdE_p = \epsilon_{\rm NT}\epsilon_{\rm dis} \dot{M}c^2$ for primary protons and $L_{e,\rm mad} = \int \dot{N}_{E_e, \rm inj} E_edE_e = (Q_e/Q_i)\epsilon_{\rm NT}\epsilon_{\rm dis} \dot{M}c^2$ for primary electrons, respectively. The cutoff energy of the particles is determined by equating the acceleration timescale with the energy loss timescale.

As the acceleration process, we consider stochastic acceleration by the MHD turbulence with a hard-sphere-like diffusion coefficient:
\begin{equation}
    D_{pp} \approx \frac{p^2}{3\eta_{\rm turb}}\left(\frac{c}{H}\right) \left(\frac{V_A}{c}\right)^2,
    \label{dif_coef}
\end{equation}
where $V_A \approx B_d/\sqrt{4 \pi \rho}$ is the Alfv\'en velocity and $B_d$ is the magnetic field in the MAD. We estimate $B_d$ as
\begin{equation}
B_d \approx \sqrt{\frac{8\pi \rho C_s^2}{\beta}} \simeq 62 ~\dot{m}_{-4}^{1/2} M_{9}^{-1/2} r_1^{-5/4} \alpha_{-0.5}^{-1/2} \beta_{-1}^{-1/2} ~ \rm G, 
\end{equation}
where $\rho \approx \dot{M}/(4\pi R_d H V_R)$ is the mass density, $H\sim (C_s/V_K)R_d$ is the scale height of the MADs, $C_s \approx V_K/2$ is the sound speed, $V_R \approx \alpha V_K$ is the radial velocity, and $V_K = \sqrt{GM_{\rm BH}/R_d} $ is the Kepler velocity \citep[see also][]{Kimura2020, Kuze+2022}. Then, we can write the acceleration timescale as 
\begin{equation}
    t_{\rm acc} = \frac{p^2}{D_{pp}}\approx 3\eta_{\rm turb}\frac{H}{c}\left(\frac{c}{V_A}\right)^2.
\end{equation}

We consider the relevant cooling processes (synchrotron, Bethe-Heitler, photomeson production), diffusive escape, and infall to the BH for the loss processes. We estimate the cooling timescales based on \citet{KimuraMurase2019}. The synchrotron cooling timescale is estimated to be 
\begin{equation}
    t_{p, \rm syn} = \frac{6\pi m_ec}{\gamma_p  \sigma_T B_d^2}\left(\frac{m_p}{m_e}\right)^3,
\end{equation} 
where $\sigma_T$ is the Thomson cross-section.

We estimate the photomeson production cooling timescale as
\begin{equation}
    t_{p\gamma}^{-1} = \frac{c}{2\gamma_p^2}\int^{\infty}_{\bar{\epsilon}_{\rm th}} d\bar{\epsilon}_\gamma \sigma_{p\gamma} \kappa_{p\gamma}\bar{\epsilon}_\gamma \int^{\infty}_{\bar{\epsilon}_\gamma/(2\gamma_p)} d\epsilon_\gamma \epsilon^{-2}_\gamma\frac{dn_\gamma}{d\epsilon_\gamma},
    \label{photomeson}
\end{equation}
where $\bar{\epsilon}_{\rm th} \simeq 145 ~\rm MeV$ is the threshold energy for photomeson production and $\bar{\epsilon}_\gamma$ is the photon energy in proton rest frame. We estimate the cooling timescale of the Bethe-Heitler process, $t_{\rm BH}$, by Equation (\ref{photomeson}) using $\sigma_{\rm BH}$ and $ \kappa_{\rm BH}$ instead of $\sigma_{p\gamma}$ and $ \kappa_{p\gamma}$, respectively. For $\sigma_{p\gamma}$ and $\kappa_{p\gamma}$, we use the fitting formulas given in \citet{MuraseNagataki2006b}, and for $\sigma_{\rm BH}$ and $\kappa_{\rm BH}$, we use those from \citet{StepneyGuilbert1983} and \citet{Chodorowski1992}, respectively. We estimate the cooling timescale as $t_{\rm cool}^{-1} = t_{p,\rm syn}^{-1}+t_{p\gamma}^{-1}+t_{\rm BH}^{-1}$.

For diffusive escape, we estimate the timescale as $t_{\rm diff} = R_d^2/D_{zz}$, where $D_{zz} = (1/3) \eta_{\rm turb} cH$ is the diffusion coefficient in space. We estimate the infall timescale as $t_{\rm fall} = R_d/V_R$, and then, the total escape timescale is given as $t_{\rm esc}^{-1} = t_{\rm fall}^{-1} + t_{\rm diff}^{-1}$.

\section{Steady-state magnetization parameter in the reconnection region}\label{base_mag}
 In this appendix, we show the detailed calculation model for the steady-state magnetization parameter in the reconnection region. We consider that the magnetic reconnection in the BH magnetosphere accelerates the particles emitting the high-energy photons, and the emitted photons interact with each other by the Breit-Wheeler process, resulting in pair production. The magnetic reconnection will produce the power-law electron energy distribution with the index $p_{\rm inj} \sim 1$ \citep[e.g.,][]{SironiSpitkovsky2014, Guo+2014, Guo+2015, Guo2016, Werner+2016}. These injected electrons rapidly cool as the cooling timescale is much shorter than the dynamical timescale \citep{Kimura+2022}, resulting in the index of nonthermal electrons, $p \sim 2$. Then, the synchrotron spectrum is $dN/dE_\gamma \propto E_\gamma^{-1.5}$ for $E_\gamma<E_{\gamma, \rm cut, ms}$, where $E_{\gamma, \rm cut, ms} = \gamma_{\rm cut, ms}^2 heB_{\rm rec}/(2\pi m_ec)$ is the photon energy emitted by the electrons with $\gamma_{\rm cut, ms}$ and $\gamma_{\rm cut, ms}$ is the cutoff Lorentz factor achieved by the magnetic reconnection in the BH magnetosphere. We assume the cutoff Lorentz factor as 
\begin{equation}
    \gamma_{\rm cut, ms} = 4\sigma_\pm.
    \label{gamcut_ms}
\end{equation} 
The synchrotron photon spectrum has an exponential cutoff for $E_{\gamma} > E_{\gamma, \rm cut, ms}$.

We consider that the produced pairs escape from the system along the magnetic field lines at the speed of light since the pair velocity ($v_{e^\pm} \simeq c$) is higher than the reconnection velocity ($\beta_{\rm rec}c \simeq 0.1c$) \citep[e.g.,][]{Guo2020}. Hence, the number density of the electron-positron pairs is estimated to be 
\begin{equation}
    n_\pm = \dot{n}_\pm \frac{l_{\rm rec}}{c},    
\end{equation} 
where $\dot{n}_\pm$ is the pair production rate. $\dot{n}_\pm$ depends on the photon spectrum, which in turn depends on the electron spectrum. Thus, $\dot{n}_\pm$ depends on $\sigma_\pm$ through Equation (\ref{gamcut_ms}). As shown in Equation (\ref{sig_std}), we estimate the steady-state magnetization parameter following the method of \citet{Chen+2022}.

\begin{figure}[tb]
    \centering
    \includegraphics[width=0.55\linewidth]{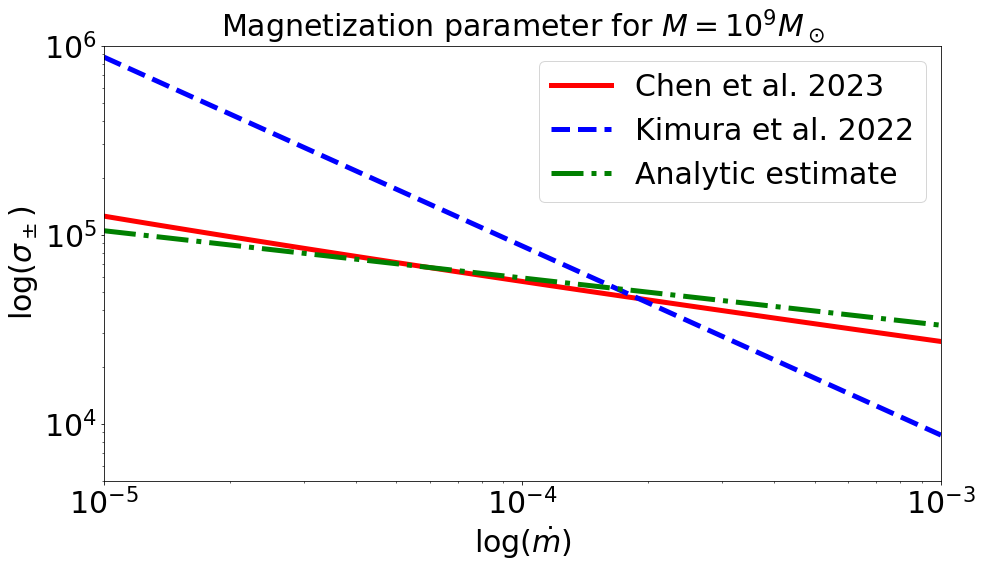}
    \caption{Magnetization parameters for BH mass of $10^9M_\odot$ as a function of $\dot{m}$, calculated in the methods of \citet{Chen+2022} (red solid line) and \citet{Kimura+2022} (blue dashed line) and estimated by Equation (\ref{sig_ana}) (green dash-dotted line).}
    \label{sig-dotm}
\end{figure}

We show $\sigma_\pm$ for the BH mass of $10^9M_\odot$ as a function of $\dot{m}$ in Figure \ref{sig-dotm}. As shown in Figure \ref{sig-dotm}, $\sigma_\pm$ approximately depends on $\dot{m}^{-1/4}$. This comes from self-regulation. For $E_{\gamma, \rm cut, ms} \gg m_ec^2$, copious pairs are produced, and then, pair loading reduces $\sigma_\pm$. On the other hand, for $E_{\gamma, \rm cut, ms} \ll m_ec^2$, plasma loading stops, and the upstream plasma will escape in the dynamical time, which will increase $\sigma_\pm$. Thus, the peak of the synchrotron spectrum is adjusted to $E_{\gamma,\rm cut, ms} \approx 0.5m_e c^2$. This can be rewritten as
\begin{equation}
\sigma_\pm \simeq 5.9\times 10^4 M_9^{1/4}\dot{m}_{-4}^{-1/4}\Phi_{\rm rec,1.2}^{-1/2}.
\label{sig_ana}
\end{equation}
\citet{Kimura+2022} also estimate the magnetization parameter after pair loading (see their Equation (11)), but our estimate of $\sigma_\pm$ differs from that given in \citet{Kimura+2022}. They assume photon spectrum with an abrupt cutoff (i.e., no photons above $E_{\gamma,\rm cut, ms}$), which leads to different values of $\sigma_\pm$ and conclusions for such a range of mass accretion rates that we consider in this paper. There is another point that differs between ours and that given in \citet{Kimura+2022}. Ours takes into account the feedback by the pair loading when $m_ec^2 < E_{\gamma,\rm cut, ms}< E_{\gamma, \rm rad}$, where $E_{\gamma,\rm rad}$ is the synchrotron photon energy for the burn-off limit (see Equation (5) in \citet{Kimura+2022}), which help achieving self-regulation given in Equation (\ref{sig_ana}). In contrast, for the cases with a mass accretion rate as low as that in Sgr A*, the estimate given in \citet{Kimura+2022} should be valid, as neither the feedback nor the tail distribution affects the pair loading process. Our model of pair loading assumes that $E_{\gamma,\rm cut, ms}$ determines the cutoff energy because of the sufficient pair loading, which contradicts the situation of low accretion rate cases where $E_{\gamma,\rm rad} < E_{\gamma,\rm cut, ms}$ is satisfied.

\section{Jet Lorenz factor and the size of the dissipation region}\label{Lorentz}
 We estimate the jet Lorentz factor and Doppler factor as follows. The very long baseline interferometry (VLBI) observations for M87 estimate the jet width as $R \approx z^{2/3}$, where $z$ is the distance from the BH \citep{Hada2013, Nakamura2018}. Thus, we assume the size of the dissipation region as $R_j = 10^2 R_g$ to have the distance of the dissipation region $z_j \simeq 10^3 R_g$. We set the jet Lorentz factor $\Gamma_j$ at the dissipation region as $\Gamma_j \beta_j \simeq 0.8$ based on the observation of the superluminal motion of the M87 jet with KaVA and VLBA \citep{Park2019}. The M87 jet inclination angle is observed as $\theta_j \approx 17 \ \rm deg$ by the observation of jet and counter jet proper motions \citep{Walker2018}. Hence, we estimate the Doppler factor to be $\delta_D \approx 1.9$.
The variability timescale is then estimated to be $t_{v} = R_j/(c \delta_D) \simeq 1.6 \times 10^6\;R_{j,2}\;{\rm s}$, which is consistent with the observed X-ray daily variability \citep{Imazawa2021, EHT_MWL2021}.
We calculate the photon spectra observed on Earth by $\nu_{\rm obs} L_{\nu,\rm obs} = \delta_D^4 \nu_{\rm src}L_{\nu, \rm src}$, where $\nu_{\rm src}L_{\nu, \rm src}$ is the intrinsic photon spectrum obtained by the jet model.

\section{Cooling and escape timescales inside the jet}\label{loss_jet}
In this appendix, we explain how to estimate the cooling and escape timescales in the jet. As explained in Section \ref{distribution}, we consider the electrons cool through synchrotron, synchrotron self-Compton, and adiabatic expansion. We estimate the cooling timescale as $t_{\rm cool}^{-1} = t_{\rm syn}^{-1} + t_{\rm SSC}^{-1} + t_{\rm ad}^{-1} $. 

We estimate the synchrotron cooling timescale to be $t_{\rm syn} = 6\pi m_ec/(\gamma \sigma_T B_{\rm dis}^2)$. We calculate the synchrotron self-Compton scattering cooling timescale in the way of subsection 2.6 of \citet{Brumenthal1970}. In this paper, we ignore the effect of the external photon field. The accretion rate is low for M87, and thus, the energy density of the external photon, such as thermal emission from a dust torus, is lower than that of the internal photon emitted by the energetic electrons. 

Since we consider a non-spherical geometry as shown in the right part of Figure \ref{Schematic_image}, the adiabatic cooling timescale differs from that usually used in previous literature. The internal energy of the emitting jet material depends on its volume, i.e., $U \propto V^{-1/3} \propto R_j^{-2/3}h_j^{-1/3}$, where $U$ is the internal energy, $V$ is the volume of the emitting jet material, and $h_j \simeq R_j$ is that of the vertical size. The pressure balance between the electromagnetic field of the jet and the ambient gas determines the jet shape to be $z \approx (R/R_g)^{3/2}R_g$. Then, we estimate the radial expansion velocity as $ \dot{R} \approx 3c/2(R/R_g)^{-1/2}$ by assuming the jet velocity as $\dot{z} \simeq c$. $\dot{R}_{j}$ is estimated to be $\dot{R}_{j} \simeq 0.1 c $ for $R_j = 100 R_g$. In contrast, the emitting jet material freely expands along the poloidal magnetic field lines, i.e., $\dot{h}_{j} \simeq c$. Thus, we estimate the adiabatic cooling timescale as $t_{\rm ad} = -U/\dot{U} \approx 3h_j/\dot{h}_{j} = 3R_j/c$.

We consider that the electrons are scattered by the KHI-induced turbulence. KHI injects the turbulence with the scale of the dissipation region, $R_j$, and cascades into the small scale. We assume that the spectrum of the turbulence is the classical Kolmogorov turbulence \citep[e.g.,][]{SanMaulik2018}. A turbulent eddy whose scale is the Larmor radius of the electrons, $r_L = \gamma m_e c^2/(eB_{\rm dis})$, scatters the electrons. We estimate the diffusion timescale as $t_{\rm esc} = R_j^2/D_R$, where $D_R \approx (4/\pi) r_L c (R_j/r_L)^{2/3}(B/\delta B)^{2} $ is the diffusion coefficient \citep[see also][]{Kulsrud2005}. 

\section{Magnetization parameter for power-law tail distribution} \label{tail_para}
Some recent PIC simulations of high $\sigma$ magnetic reconnection exhibit a power-law tail at $\gamma > \sigma$ in the nonthermal particle spectrum \citep{Hakobyan2021, Hakobyan2023}. We estimate the magnetization parameter at the jet base for the power-law tail distribution in a similar manner to Appendix D in \citet{Kimura+2022}. For simplicity, we assume the nonthermal particle spectrum as 
\begin{equation}
      \dot{N}_{e,\rm inj} = \dot{N}_0
    \begin{cases}
        \left(\frac{\gamma}{4\sigma} \right) ^{-p} \ (\gamma < 4\sigma  ) \\
        \left(\frac{\gamma}{4\sigma} \right)^{-p_{\rm tail}} \ (4\sigma < \gamma < \gamma_{\rm rad})\\
    \end{cases}.
\end{equation}
Here, we set $p < 2$ and $2< p_{\rm tail} \leq 3$. We estimate the normalization, $\dot{N}_0$, by equating the electron luminosity with the energy release rate by the magnetic reconnection $L_{\rm rec}$, i.e., $L_{\rm rec} \approx L_e = \int \dot{N}_{\rm inj} \gamma m_ec^2 d\gamma$. We estimate the reconnection luminosity as 
\begin{align}
    L_{\rm rec} &\approx 2 l_{\rm rec}^2 \frac{B_{\rm rec}^2}{8\pi}\beta_{\rm rec} c \\
    &\simeq 7.9 \times 10^{41} M_9 \dot{m}_{-4}\beta_{\rm rec, -1} \Phi_{\rm rec, 1.2}^2 \ \rm erg\ s^{-1}.
\end{align}
We write the electron luminosity at $\gamma = 4\sigma$ as $L_{\rm max}$ and estimate $L_{\rm max}$ as   
\begin{align}
    L_{\rm max} = \frac{L_{\rm rec}}{\frac{1}{2-p} \left[ 1- \left( \frac{E_\sigma}{m_ec^2}\right)^{-p +2}\right] + \frac{1}{2-p_{\rm tail}} \left[\left(\frac{4\sigma}{\gamma_{\rm rad}}\right)^{p_{\rm tail}-2} - 1\right] },
\end{align}
where $E_\sigma = 4\sigma m_ec^2$. Since the system is in a fast cooling regime, all the electron energy is converted to the photons via synchrotron radiation. Then, we write the synchrotron spectrum as 
\begin{equation}
      E_\gamma L_{E_\gamma} \approx L_{\rm max}
    \begin{cases}
        \left(\frac{E_\gamma}{E_{\gamma, \sigma}} \right) ^{(2-p)/2} \ (E_\gamma < E_{\gamma, \sigma}  ) \\
        \left(\frac{E_\gamma}{E_{\gamma, \sigma}} \right)^{(2-p_{\rm tail})/2} \ (E_{\gamma, \sigma} < E_\gamma < E_{\gamma, \gamma_{\rm rad}})\\
    \end{cases},
\end{equation}
where $E_{\gamma, \sigma} = (4\sigma)^2 heB_{\rm rec}/(2\pi m_ec)$  and $E_{\gamma,\rm rad} = \gamma_{\rm rad}^2 heB_{\rm rec}/(2\pi m_ec)$ are the synchrotron photons emitted by the electrons with $4\sigma$ and $\gamma_{\rm rad}$, respectively. 

Due to the tail distribution, the synchrotron spectrum has two components, which forces us to consider three cases depending on the relation among $E_{\gamma, \sigma}$, $E_{\gamma, \rm rad}$, and $E_{\rm thr} = 2m_e c^2$. Hereafter, we consider interactions of $E_{\gamma,1}$ with $E_{\gamma,2} = (2m_ec^2)^2/E_{\gamma,1}$ and the energy range of  $E_{\gamma,1} > E_{\rm thr}$. 

First, we consider $E_{\gamma, \sigma} < (2m_ec^2)/E_{\gamma, {\rm rad}}$. In this case, both $E_{\gamma,1}$ and $E_{\gamma,2}$ exist in the synchrotron spectrum with $(2-p_{\rm tail}/2)$. We integrate for $E_{\gamma,1}$ to estimate the total pair number density, which is estimated to be
\begin{align}
    n_{\pm, \rm tot} &= \int_{E_{\rm thr}}^{E_{\gamma, \rm rad}} \frac{3L_{\rm max}^2 \sigma_{\gamma\gamma} }{4\pi l_{\rm rec}^3 c^2 E_{\gamma,\sigma}^2}\left( \frac{E_{\gamma, \sigma}}{2m_ec^2}\right)^{p_{\rm tail}} d\log E_{\gamma,1} \\
    &=  \frac{3L_{\rm max}^2 \sigma_{\gamma\gamma} }{4\pi l_{\rm rec}^3 c^2 E_{\gamma,\sigma}^2}\left( \frac{E_{\gamma, \sigma}}{2m_ec^2}\right)^{p_{\rm tail}} \log \left( \frac{E_{\gamma, \rm rad}}{E_{\rm thr}}\right), 
    \label{A;1}
\end{align}
where $\sigma_{\gamma\gamma}\approx f_{\gamma\gamma}\sigma_T$ is the approximated cross section for the Breit-Wheeler process and $f_{\gamma\gamma} \sim 0.2$. 

Second, we consider $(2m_ec^2)/(E_{\gamma, \rm rad})< E_{\gamma, \sigma} < E_{\rm thr}$. In this case, $E_{\gamma,2}$ exists in both components whose index is $(2-p)/2$ for $E_{\gamma,1} > E_{\gamma,c} = (2m_ec^2)/(E_{\gamma, \sigma})$ and is $(2-p_{\rm tail})/2$ for $E_{\gamma,1} < E_{\gamma,c}$. Then, we estimate the total pair number density to be
\begin{align}
    n_{\pm, \rm tot} &=  \frac{3L_{\rm max}^2 \sigma_{\gamma\gamma} }{4\pi l_{\rm rec}^3 c^2} \left[
    \int_{E_{\rm thr}}^{E_{\gamma, c}} \frac{1}{E_{\gamma,\sigma}^2}\left( \frac{E_{\gamma, \sigma}}{2m_ec^2}\right)^{p_{\rm tail}} d\log E_{\gamma,1} +
    \int_{E_{\gamma,c}}^{E_{\gamma, \rm rad}} E_{\gamma, \sigma}^{(p+p_{\rm tail})/2 -2} (2m_ec^2)^{-p} E_{\gamma,1}^{(p-p_{\rm tail})/2}  d\log E_{\gamma,1}  \right] \\
    &=  \frac{3L_{\rm max}^2 \sigma_{\gamma\gamma} }{4\pi l_{\rm rec}^3 c^2 }
    \left[\frac{1}{E_{\gamma,\sigma}^2}\left( \frac{E_{\gamma, \sigma}}{2m_ec^2}\right)^{p_{\rm tail}} \log \left( \frac{E_{\gamma, \rm rad}}{E_{\rm thr}}\right)
    +E_{\gamma, \sigma}^{(p+p_{\rm tail})/2-2} (2m_ec^2)^{-p}\frac{2}{p-p_{\rm tail}} \left( E_{\gamma, \rm rad}^{(p-p_{\rm tail})/2} - E_{\gamma,c}^{(p-p_{\rm tail})/2}\right)
    \right].
    \label{A;2}
\end{align}

Third, we consider $E_{\gamma,\sigma} > E_{\rm thr}$. In this case, $E_{\gamma,1}$ exists in both components whose index is $(2-p)/2$ and $(2-p_{\rm tail})/2$, and $E_{\gamma,2}$ is in the synchrotron spectrum whose index is $(2-p)/2$. We estimate the total pair number density to be
\begin{align}
    n_{\pm, \rm tot} &=  \frac{3L_{\rm max}^2 \sigma_{\gamma\gamma} }{4\pi l_{\rm rec}^3 c^2 } \left[
    \int_{E_{\rm thr}}^{E_{\gamma, \sigma}} \frac{1}{(2m_ec^2)^3}\left( \frac{E_{\gamma, \sigma}}{2m_ec^2}\right)^{p-2} d\log E_{\gamma,1} +
    \int_{E_{\gamma,\sigma}}^{E_{\gamma, \rm rad}}E_{\gamma, \sigma}^{(p+p_{\rm tail})/2 -2} (2m_ec^2)^{-p} E_{\gamma,1}^{(p-p_{\rm tail})/2}d\log E_{\gamma,1}  \right] \\
    &=  \frac{3L_{\rm max}^2 \sigma_{\gamma\gamma} }{4\pi l_{\rm rec}^3 c^2}
    \left[\frac{1}{(2m_ec^2)^3}\left( \frac{E_{\gamma, \sigma}}{2m_ec^2}\right)^{p-2} \log \left( \frac{E_{\gamma, \sigma}}{E_{\rm thr}}\right)
    + E_{\gamma, \sigma}^{(p+p_{\rm tail})/2-2} (2m_ec^2)^{-p}\frac{2}{p-p_{\rm tail}} \left( E_{\gamma, \rm rad}^{(p-p_{\rm tail})/2} - E_{\gamma,\sigma}^{(p-p_{\rm tail})/2}\right)
    \right].
    \label{A;3}
\end{align}

\begin{figure}
    \centering
   \gridline{\fig{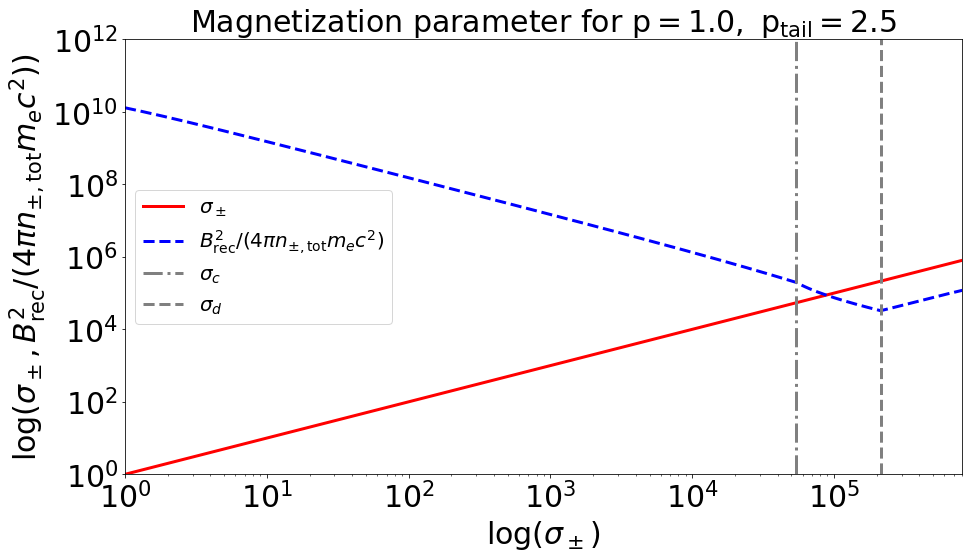}{0.55\linewidth}{}}
    \caption{Magnetization parameters for M87 with $p = 1.0, ~ p_{\rm tail} = 2.5$. The red line is $\sigma_\pm$ and the blue dashed line is $B_{\rm rec}^2/(4\pi n_{\pm, \rm tot} m_ec^2)$, respectively. The vertical gray dot-dashed line and dashed line are magnetization parameters of $E_{\gamma,\sigma} = (2m_ec^2)/E_{\gamma, \rm rad}$ and $E_{\gamma, \sigma} = E_{\rm thr}$, respectively.}
    \label{fig:mp}
\end{figure}

Since the pair number density depends on the magnetization parameter, we calculate the magnetization parameter by solving the $\sigma_\pm = B_{\rm rec}^2/(4\pi n_{\pm, \rm tot} m_ec^2)$ with Equations (\ref{A;1}),(\ref{A;2}), and (\ref{A;3}). We show $\sigma_\pm$ and $B_{\rm rec}^2/(4\pi n_{\pm, \rm tot} m_ec^2)$ in Figure \ref{fig:mp} as the red line and blue dashed line, respectively. We estimate the magnetization parameter as $\sigma_\pm \simeq8.8 \times10^{4}$ and $1.0 \times10^{5}$ for $p=1.0, ~p_{\rm tail} = 2.5$ and $p=1.0, ~p_{\rm tail} = 3.0$. As shown in Figure \ref{fig:enter-label}, the estimated magnetization parameter is almost the same as estimated in the way of \cite{Chen+2022} and hardly changes its value with the tail index due to self-regulation of the synchrotron spectrum to peak at $E_\gamma \sim m_ec^2$. 

\bibliography{sample631}{}

\begin{thebibliography}{}
\expandafter\ifx\csname natexlab\endcsname\relax\def\natexlab#1{#1}\fi
\providecommand{\url}[1]{\href{#1}{#1}}
\providecommand{\dodoi}[1]{doi:~\href{http://doi.org/#1}{\nolinkurl{#1}}}
\providecommand{\doeprint}[1]{\href{http://ascl.net/#1}{\nolinkurl{http://ascl.net/#1}}}
\providecommand{\doarXiv}[1]{\href{https://arxiv.org/abs/#1}{\nolinkurl{https://arxiv.org/abs/#1}}}

\bibitem[{{Aartsen} {et~al.}(2021){Aartsen}, {Abbasi}, {Ackermann}, {Adams}, {Aguilar}, {Ahlers}, {Ahrens}, {Alispach}, {Allison}, {Amin}, {Andeen}, {Anderson}, {Ansseau}, {Anton}, {Arg{\"u}elles}, {Arlen}, {Auffenberg}, {Axani}, {Bagherpour}, {Bai}, {Balagopal V}, {Barbano}, {Bartos}, {Bastian}, {Basu}, {Baum}, {Baur}, {Bay}, {Beatty}, {Becker}, {Tjus}, {BenZvi}, {Berley}, {Bernardini}, {Besson}, {Binder}, {Bindig}, {Blaufuss}, {Blot}, {Bohm}, {Bohmer}, {B{\"o}ser}, {Botner}, {B{\"o}ttcher}, {Bourbeau}, {Bourbeau}, {Bradascio}, {Braun}, {Bron}, {Brostean-Kaiser}, {Burgman}, {Burley}, {Buscher}, {Busse}, {Bustamante}, {Campana}, {Carnie-Bronca}, {Carver}, {Chen}, {Chen}, {Cheung}, {Chirkin}, {Choi}, {Clark}, {Clark}, {Classen}, {Coleman}, {Collin}, {Connolly}, {Conrad}, {Coppin}, {Correa}, {Cowen}, {Cross}, {Dave}, {Deaconu}, {De Clercq}, {DeLaunay}, {De Kockere}, {Dembinski}, {Deoskar}, {De Ridder}, {Desai}, {Desiati}, {de Vries}, {de Wasseige}, {de With}, {DeYoung}, {Dharani}, {Diaz}, {D{\'\i}az-V{\'e}lez},
  {Dujmovic}, {Dunkman}, {DuVernois}, {Dvorak}, {Ehrhardt}, {Eller}, {Engel}, {Evans}, {Evenson}, {Fahey}, {Farrag}, {Fazely}, {Felde}, {Fienberg}, {Filimonov}, {Finley}, {Fischer}, {Fox}, {Franckowiak}, {Friedman}, {Fritz}, {Gaisser}, {Gallagher}, {Ganster}, {Garcia-Fernandez}, {Garrappa}, {Gartner}, {Gerhard}, {Gernhaeuser}, {Ghadimi}, {Glaser}, {Glauch}, {Gl{\"u}senkamp}, {Goldschmidt}, {Gonzalez}, {Goswami}, {Grant}, {Gr{\'e}goire}, {Griffith}, {Griswold}, {G{\"u}nd{\"u}z}, {Haack}, {Hallgren}, {Halliday}, {Halve}, {Halzen}, {Hanson}, {Hanson}, {Hardin}, {Haugen}, {Haungs}, {Hauser}, {Hebecker}, {Heinen}, {Heix}, {Helbing}, {Hellauer}, {Henningsen}, {Hickford}, {Hignight}, {Hill}, {Hill}, {Hoffman}, {Hoffmann}, {Hoffmann}, {Hoinka}, {Hokanson-Fasig}, {Holzapfel}, {Hoshina}, {Huang}, {Huber}, {Huber}, {Huege}, {Hughes}, {Hultqvist}, {H{\"u}nnefeld}, {Hussain}, {In}, {Iovine}, {Ishihara}, {Jansson}, {Japaridze}, {Jeong}, {Jones}, {Jonske}, {Joppe}, {Kalekin}, {Kang}, {Kang}, {Kang}, {Kappes}, {Kappesser},
  {Karg}, {Karl}, {Karle}, {Katori}, {Katz}, {Kauer}, {Keivani}, {Kellermann}, {Kelley}, {Kheirandish}, {Kim}, {Kin}, {Kintscher}, {Kiryluk}, {Kittler}, {Kleifges}, {Klein}, {Koirala}, {Kolanoski}, {K{\"o}pke}, {Kopper}, {Kopper}, {Koskinen}, {Koundal}, {Kovacevich}, {Kowalski}, {Krauss}, {Krings}, {Kr{\"u}ckl}, {Kulacz}, {Kurahashi}, {Gualda}, {Lahmann}, {Lanfranchi}, {Larson}, {Latif}, {Lauber}, {Lazar}, {Leonard}, {Leszczy{\'n}ska}, {Li}, {Liu}, {Lohfink}, {LoSecco}, {Mariscal}, {Lu}, {Lucarelli}, {Ludwig}, {L{\"u}nemann}, {Luszczak}, {Lyu}, {Ma}, {Madsen}, {Maggi}, {Mahn}, {Makino}, {Mallik}, {Mancina}, {Mandalia}, {Mari{\c{s}}}, {Marka}, {Marka}, {Maruyama}, {Mase}, {Maunu}, {McNally}, {Meagher}, {Medina}, {Meier}, {Meighen-Berger}, {Merz}, {Meyers}, {Micallef}, {Mockler}, {Moment{\'e}}, {Montaruli}, {Moore}, {Morse}, {Moulai}, {Muth}, {Naab}, {Nagai}, {Nam}, {Nauman}, {Necker}, {Neer}, {Nelles}, {Nguyễn}, {Niederhausen}, {Nisa}, {Nowicki}, {Nygren}, {Oberla}, {Pollmann}, {Oehler}, {Olivas},
  {O'Sullivan}, {Pan}, {Pandya}, {Pankova}, {Papp}, {Park}, {Parker}, {Paudel}, {Peiffer}, {P{\'e}rez de los Heros}, {Petersen}, {Philippen}, {Pieloth}, {Pieper}, {Pinfold}, {Pizzuto}, {Plaisier}, {Plum}, {Popovych}, {Porcelli}, {Rodriguez}, {Price}, {Przybylski}, {Raab}, {Raissi}, {Rameez}, {Rauch}, {Rawlins}, {Rea}, {Rehman}, {Reimann}, {Renschler}, {Renzi}, {Resconi}, {Reusch}, {Rhode}, {Richman}, {Riedel}, {Riegel}, {Roberts}, {Robertson}, {Roellinghoff}, {Rongen}, {Rott}, {Ruhe}, {Ryckbosch}, {Cantu}, {Safa}, {Herrera}, {Sandrock}, {Sandroos}, {Sandstrom}, {Santander}, {Sarkar}, {Sarkar}, {Satalecka}, {Scharf}, {Schaufel}, {Schieler}, {Schlunder}, {Schmidt}, {Schneider}, {Schneider}, {Schr{\"o}der}, {Schumacher}, {Sclafani}, {Seckel}, {Seunarine}, {Shaevitz}, {Sharma}, {Shefali}, {Silva}, {Smith}, {Smithers}, {Snihur}, {Soedingrekso}, {Soldin}, {S{\"o}ldner-Rembold}, {Song}, {Southall}, {Spiczak}, {Spiering}, {Stachurska}, {Stamatikos}, {Stanev}, {Stein}, {Stettner}, {Steuer}, {Stezelberger}, {Stokstad},
  {Strotjohann}, {St{\"u}rwald}, {Stuttard}, {Sullivan}, {Taboada}, {Taketa}, {Tanaka}, {Tenholt}, {Ter-Antonyan}, {Terliuk}, {Tilav}, {Tollefson}, {Tomankova}, {T{\"o}nnis}, {Torres}, {Toscano}, {Tosi}, {Trettin}, {Tselengidou}, {Tung}, {Turcati}, {Turcotte}, {Turley}, {Twagirayezu}, {Ty}, {Unger}, {Elorrieta}, {Vandenbroucke}, {van Eijk}, {van Eijndhoven}, {Vannerom}, {van Santen}, {Veberic}, {Verpoest}, {Vieregg}, {Vraeghe}, {Walck}, {Watson}, {Weaver}, {Weindl}, {Weinstock}, {Weiss}, {Weldert}, {Welling}, {Wendt}, {Werthebach}, {Whitehorn}, {Wiebe}, {Wiebusch}, {Williams}, {Wissel}, {Wolf}, {Wood}, {Woschnagg}, {Wrede}, {Wren}, {Wulff}, {Xu}, {Xu}, {Yanez}, {Yoshida}, {Yuan}, {Zhang}, {Zierke}, \& {Z{\"o}cklein}}]{IceCubeGen22021}
{Aartsen}, M.~G., {Abbasi}, R., {Ackermann}, M., {et~al.} 2021, Journal of Physics G Nuclear Physics, 48, 060501, \dodoi{10.1088/1361-6471/abbd48}

\bibitem[{{Abdo} {et~al.}(2009){Abdo}, {Ackermann}, {Ajello}, {Atwood}, {Axelsson}, {Baldini}, {Ballet}, {Barbiellini}, {Bastieri}, {Bechtol}, {Bellazzini}, {Berenji}, {Blandford}, {Bloom}, {Bonamente}, {Borgland}, {Bregeon}, {Brez}, {Brigida}, {Bruel}, {Burnett}, {Caliandro}, {Cameron}, {Cannon}, {Caraveo}, {Casandjian}, {Cavazzuti}, {Cecchi}, {{\c{C}}elik}, {Charles}, {Cheung}, {Chiang}, {Ciprini}, {Claus}, {Cohen-Tanugi}, {Colafrancesco}, {Conrad}, {Costamante}, {Cutini}, {Davis}, {Dermer}, {de Angelis}, {de Palma}, {Digel}, {Donato}, {Silva}, {Drell}, {Dubois}, {Dumora}, {Edmonds}, {Farnier}, {Favuzzi}, {Fegan}, {Finke}, {Focke}, {Fortin}, {Frailis}, {Fukazawa}, {Funk}, {Fusco}, {Gargano}, {Gasparrini}, {Gehrels}, {Georganopoulos}, {Germani}, {Giebels}, {Giglietto}, {Giommi}, {Giordano}, {Giroletti}, {Glanzman}, {Godfrey}, {Grenier}, {Grondin}, {Grove}, {Guillemot}, {Guiriec}, {Hanabata}, {Harding}, {Hayashida}, {Hays}, {Horan}, {J{\'o}hannesson}, {Johnson}, {Johnson}, {Johnson}, {Johnson}, {Kamae},
  {Katagiri}, {Kataoka}, {Kawai}, {Kerr}, {Kn{\"o}dlseder}, {Kocian}, {Kuss}, {Lande}, {Latronico}, {Lemoine-Goumard}, {Longo}, {Loparco}, {Lott}, {Lovellette}, {Lubrano}, {Madejski}, {Makeev}, {Mazziotta}, {McConville}, {McEnery}, {Meurer}, {Michelson}, {Mitthumsiri}, {Mizuno}, {Moiseev}, {Monte}, {Monzani}, {Morselli}, {Moskalenko}, {Murgia}, {Nolan}, {Norris}, {Nuss}, {Ohsugi}, {Omodei}, {Orlando}, {Ormes}, {Ozaki}, {Paneque}, {Panetta}, {Parent}, {Pelassa}, {Pepe}, {Pesce-Rollins}, {Piron}, {Porter}, {Rain{\`o}}, {Rando}, {Razzano}, {Reimer}, {Reimer}, {Reposeur}, {Ritz}, {Rochester}, {Rodriguez}, {Romani}, {Roth}, {Ryde}, {Sadrozinski}, {Sambruna}, {Sanchez}, {Sander}, {Saz Parkinson}, {Scargle}, {Sgr{\`o}}, {Shaw}, {Smith}, {Smith}, {Spandre}, {Spinelli}, {Strickman}, {Suson}, {Tajima}, {Takahashi}, {Tanaka}, {Taylor}, {Thayer}, {Thompson}, {Tibaldo}, {Torres}, {Tosti}, {Tramacere}, {Uchiyama}, {Usher}, {Vasileiou}, {Vilchez}, {Waite}, {Wang}, {Winer}, {Wood}, {Ylinen}, {Ziegler}, {Harris}, {Massaro},
  \& {Stawarz}}]{Abdo+2009}
{Abdo}, A.~A., {Ackermann}, M., {Ajello}, M., {et~al.} 2009, \apj, 707, 55, \dodoi{10.1088/0004-637X/707/1/55}

\bibitem[{{Actis} {et~al.}(2011){Actis}, {Agnetta}, {Aharonian}, {Akhperjanian}, {Aleksi{\'c}}, {Aliu}, {Allan}, {Allekotte}, {Antico}, {Antonelli}, {Antoranz}, {Aravantinos}, {Arlen}, {Arnaldi}, {Artmann}, {Asano}, {Asorey}, {B{\"a}hr}, {Bais}, {Baixeras}, {Bajtlik}, {Balis}, {Bamba}, {Barbier}, {Barcel{\'o}}, {Barnacka}, {Barnstedt}, {Barres de Almeida}, {Barrio}, {Basso}, {Bastieri}, {Bauer}, {Becerra}, {Becherini}, {Bechtol}, {Becker}, {Beckmann}, {Bednarek}, {Behera}, {Beilicke}, {Belluso}, {Benallou}, {Benbow}, {Berdugo}, {Berger}, {Bernardino}, {Bernl{\"o}hr}, {Biland}, {Billotta}, {Bird}, {Birsin}, {Bissaldi}, {Blake}, {Blanch}, {Bobkov}, {Bogacz}, {Bogdan}, {Boisson}, {Boix}, {Bolmont}, {Bonanno}, {Bonardi}, {Bonev}, {Borkowski}, {Botner}, {Bottani}, {Bourgeat}, {Boutonnet}, {Bouvier}, {Brau-Nogu{\'e}}, {Braun}, {Bretz}, {Briggs}, {Brun}, {Brunetti}, {Buckley}, {Bugaev}, {B{\"u}hler}, {Bulik}, {Busetto}, {Buson}, {Byrum}, {Cailles}, {Cameron}, {Canestrari}, {Cantu}, {Carmona}, {Carosi}, {Carr},
  {Carton}, {Casiraghi}, {Castarede}, {Catalano}, {Cavazzani}, {Cazaux}, {Cerruti}, {Cerruti}, {Chadwick}, {Chiang}, {Chikawa}, {Cie{\'s}lar}, {Ciesielska}, {Cillis}, {Clerc}, {Colin}, {Colom{\'e}}, {Compin}, {Conconi}, {Connaughton}, {Conrad}, {Contreras}, {Coppi}, {Corlier}, {Corona}, {Corpace}, {Corti}, {Cortina}, {Costantini}, {Cotter}, {Courty}, {Couturier}, {Covino}, {Croston}, {Cusumano}, {Daniel}, {Dazzi}, {de Angelis}, {de Cea Del Pozo}, {de Gouveia Dal Pino}, {de Jager}, {de La Calle P{\'e}rez}, {de La Vega}, {de Lotto}, {de Naurois}, {de O{\~n}a Wilhelmi}, {de Souza}, {Decerprit}, {Deil}, {Delagnes}, {Deleglise}, {Delgado}, {Dettlaff}, {di Paolo}, {di Pierro}, {D{\'\i}az}, {Dick}, {Dickinson}, {Digel}, {Dimitrov}, {Disset}, {Djannati-Ata{\"\i}}, {Doert}, {Domainko}, {Dorner}, {Doro}, {Dournaux}, {Dravins}, {Drury}, {Dubois}, {Dubois}, {Dubus}, {Dufour}, {Durand}, {Dyks}, {Dyrda}, {Edy}, {Egberts}, {Eleftheriadis}, {Elles}, {Emmanoulopoulos}, {Enomoto}, {Ernenwein}, {Errando}, {Etchegoyen},
  {Falcone}, {Farakos}, {Farnier}, {Federici}, {Feinstein}, {Ferenc}, {Fillin-Martino}, {Fink}, {Finley}, {Finley}, {Firpo}, {Florin}, {F{\"o}hr}, {Fokitis}, {Font}, {Fontaine}, {Fontana}, {F{\"o}rster}, {Fortson}, {Fouque}, {Fransson}, {Fraser}, {Fresnillo}, {Fruck}, {Fujita}, {Fukazawa}, {Funk}, {G{\"a}bele}, {Gabici}, {Gadola}, {Galante}, {Gallant}, {Garc{\'\i}a}, {Garc{\'\i}a L{\'o}pez}, {Garrido}, {Garrido}, {Gasc{\'o}n}, {Gasq}, {Gaug}, {Gaweda}, {Geffroy}, {Ghag}, {Ghedina}, {Ghigo}, {Gianakaki}, {Giarrusso}, {Giavitto}, {Giebels}, {Giro}, {Giubilato}, {Glanzman}, {Glicenstein}, {Gochna}, {Golev}, {G{\'o}mez Berisso}, {Gonz{\'a}lez}, {Gonz{\'a}lez}, {Gra{\~n}ena}, {Graciani}, {Granot}, {Gredig}, {Green}, {Greenshaw}, {Grimm}, {Grube}, {Grudzi{\'n}ska}, {Grygorczuk}, {Guarino}, {Guglielmi}, {Guilloux}, {Gunji}, {Gyuk}, {Hadasch}, {Haefner}, {Hagiwara}, {Hahn}, {Hallgren}, {Hara}, {Hardcastle}, {Hassan}, {Haubold}, {Hauser}, {Hayashida}, {Heller}, {Henri}, {Hermann}, {Herrero}, {Hinton}, {Hoffmann},
  {Hofmann}, {Hofverberg}, {Horns}, {Hrupec}, {Huan}, {Huber}, {Huet}, {Hughes}, {Hultquist}, {Humensky}, {Huppert}, {Ibarra}, {Illa}, {Ingjald}, {Inoue}, {Inoue}, {Ioka}, {Jablonski}, {Jacholkowska}, {Janiak}, {Jean}, {Jensen}, {Jogler}, {Jung}, {Kaaret}, {Kabuki}, {Kakuwa}, {Kalkuhl}, {Kankanyan}, {Kapala}, {Karastergiou}, {Karczewski}, {Karkar}, {Karlsson}, {Kasperek}, {Katagiri}, {Katarzy{\'n}ski}, {Kawanaka}, {K{\c{e}}dziora}, {Kendziorra}, {Kh{\'e}lifi}, {Kieda}, {Kifune}, {Kihm}, {Klepser}, {Klu{\'z}niak}, {Knapp}, {Knappy}, {Kneiske}, {Kn{\"o}dlseder}, {K{\"o}ck}, {Kodani}, {Kohri}, {Kokkotas}, {Komin}, {Konopelko}, {Kosack}, {Kossakowski}, {Kostka}, {Kotu{\l}a}, {Kowal}, {Kozio{\l}}, {Kr{\"a}henb{\"u}hl}, {Krause}, {Krawczynski}, {Krennrich}, {Kretzschmann}, {Kubo}, {Kudryavtsev}, {Kushida}, {La Barbera}, {La Parola}, {La Rosa}, {L{\'o}pez}, {Lamanna}, {Laporte}, {Lavalley}, {Le Flour}, {Le Padellec}, {Lenain}, {Lessio}, {Lieunard}, {Lindfors}, {Liolios}, {Lohse}, {Lombardi}, {Lopatin}, {Lorenz},
  {Lubi{\'n}ski}, {Luz}, {Lyard}, {Maccarone}, {Maccarone}, {Maier}, {Majumdar}, {Maltezos}, {Ma{\l}kiewicz}, {Ma{\~n}{\'a}}, {Manalaysay}, {Maneva}, {Mangano}, {Manigot}, {Mar{\'\i}n}, {Mariotti}, {Markoff}, {Mart{\'\i}nez}, {Mart{\'\i}nez}, {Mastichiadis}, {Matsumoto}, {Mattiazzo}, {Mazin}, {McComb}, {McCubbin}, {McHardy}, {Medina}, {Melkumyan}, {Mendes}, {Mertsch}, {Meucci}, {Micha{\l}owski}, {Micolon}, {Mineo}, {Mirabal}, {Mirabel}, {Miranda}, {Mirzoyan}, {Mizuno}, {Moal}, {Moderski}, {Molinari}, {Monteiro}, {Moralejo}, {Morello}, {Mori}, {Motta}, {Mottez}, {Moulin}, {Mukherjee}, {Munar}, {Muraishi}, {Murase}, {Murphy}, {Nagataki}, {Naito}, {Nakamori}, {Nakayama}, {Naumann}, {Naumann}, {Nayman}, {Nedbal}, {Nied{\'z}wiecki}, {Niemiec}, {Nikolaidis}, {Nishijima}, {Nolan}, {Nowak}, {O'Brien}, {Ochoa}, {Ohira}, {Ohishi}, {Ohka}, {Okumura}, {Olivetto}, {Ong}, {Orito}, {Orr}, {Osborne}, {Ostrowski}, {Otero}, {Otte}, {Ovcharov}, {Oya}, {Ozi{\c{e}}b{\l}o}, {Paiano}, {Pallota}, {Panazol}, {Paneque}, {Panter},
  {Paoletti}, {Papyan}, {Paredes}, {Pareschi}, {Parsons}, {Paz Arribas}, {Pedaletti}, {Pepato}, {Persic}, {Petrucci}, {Peyaud}, {Piechocki}, {Pita}, {Pivato}, {P{\l}atos}, {Platzer}, {Pogosyan}, {Pohl}, {Pojma{\'n}ski}, {Ponz}, {Potter}, {Prandini}, {Preece}, {Prokoph}, {P{\"u}hlhofer}, {Punch}, {Quel}, {Quirrenbach}, {Rajda}, {Rando}, {Rataj}, {Raue}, {Reimann}, {Reimann}, {Reimer}, {Reimer}, {Renaud}, {Renner}, {Reymond}, {Rhode}, {Rib{\'o}}, {Ribordy}, {Rico}, {Rieger}, {Ringegni}, {Ripken}, {Ristori}, {Rivoire}, {Rob}, {Rodriguez}, {Roeser}, {Romano}, {Romero}, {Rosier-Lees}, {Rovero}, {Roy}, {Royer}, {Rudak}, {Rulten}, {Ruppel}, {Russo}, {Ryde}, {Sacco}, {Saggion}, {Sahakian}, {Saito}, {Saito}, {Sakaki}, {Salazar}, {Salini}, {S{\'a}nchez}, {S{\'a}nchez Conde}, {Santangelo}, {Santos}, {Sanuy}, {Sapozhnikov}, {Sarkar}, {Scalzotto}, {Scapin}, {Scarcioffolo}, {Schanz}, {Schlenstedt}, {Schlickeiser}, {Schmidt}, {Schmoll}, {Schroedter}, {Schultz}, {Schultze}, {Schulz}, {Schwanke}, {Schwarzburg}, {Schweizer},
  {Seiradakis}, {Selmane}, {Seweryn}, {Shayduk}, {Shellard}, {Shibata}, {Sikora}, {Silk}, {Sillanp{\"a}{\"a}}, {Sitarek}, {Skole}, {Smith}, {Sobczy{\'n}ska}, {Sofo Haro}, {Sol}, {Spanier}, {Spiga}, {Spyrou}, {Stamatescu}, {Stamerra}, {Starling}, {Stawarz}, {Steenkamp}, {Stegmann}, {Steiner}, {Stergioulas}, {Sternberger}, {Stinzing}, {Stodulski}, {Straumann}, {Su{\'a}rez}, {Suchenek}, {Sugawara}, {Sulanke}, {Sun}, {Supanitsky}, {Sutcliffe}, {Szanecki}, {Szepieniec}, {Szostek}, {Szymkowiak}, {Tagliaferri}, {Tajima}, {Takahashi}, {Takahashi}, {Takalo}, {Takami}, {Talbot}, {Tam}, {Tanaka}, {Tanimori}, {Tavani}, {Tavernet}, {Tchernin}, {Tejedor}, {Telezhinsky}, {Temnikov}, {Tenzer}, {Terada}, {Terrier}, {Teshima}, {Testa}, {Tibaldo}, {Tibolla}, {Tluczykont}, {Todero Peixoto}, {Tokanai}, {Tokarz}, {Toma}, {Torres}, {Tosti}, {Totani}, {Toussenel}, {Vallania}, {Vallejo}, {van der Walt}, {van Eldik}, {Vandenbroucke}, {Vankov}, {Vasileiadis}, {Vassiliev}, {Vegas}, {Venter}, {Vercellone}, {Veyssiere}, {Vialle},
  {Videla}, {Vincent}, {Vink}, {Vlahakis}, {Vlahos}, {Vogler}, {Vollhardt}, {Volpe}, {von Gunten}, {Vorobiov}, {Wagner}, {Wagner}, {Wagner}, {Wakely}, {Walter}, {Walter}, {Warwick}, {Wawer}, {Wawrzaszek}, {Webb}, {Wegner}, {Weinstein}, {Weitzel}, {Welsing}, {Wetteskind}, {White}, {Wierzcholska}, {Wilkinson}, {Williams}, {Winde}, {Wischnewski}, {Wi{\'s}niewski}, {Wolczko}, {Wood}, {Xiong}, {Yamamoto}, {Yamaoka}, {Yamazaki}, {Yanagita}, {Yoffo}, {Yonetani}, {Yoshida}, {Yoshida}, {Yoshikoshi}, {Zabalza}, {Zagda{\'n}ski}, {Zajczyk}, {Zdziarski}, {Zech}, {Zi{\c{e}}tara}, {Zi{\'o}{\l}kowski}, {Zitelli}, \& {Zychowski}}]{CTA2011}
{Actis}, M., {Agnetta}, G., {Aharonian}, F., {et~al.} 2011, Experimental Astronomy, 32, 193, \dodoi{10.1007/s10686-011-9247-0}

\bibitem[{{Aguilar} {et~al.}(2021){Aguilar}, {Allison}, {Beatty}, {Bernhoff}, {Besson}, {Bingefors}, {Botner}, {Buitink}, {Carter}, {Clark}, {Connolly}, {Dasgupta}, {de Kockere}, {de Vries}, {Deaconu}, {DuVernois}, {Feigl}, {Garc{\'\i}a-Fern{\'a}ndez}, {Glaser}, {Hallgren}, {Hallmann}, {Hanson}, {Hendricks}, {Hokanson-Fasig}, {Hornhuber}, {Hughes}, {Karle}, {Kelley}, {Klein}, {Krebs}, {Lahmann}, {Magnuson}, {Meures}, {Meyers}, {Nelles}, {Novikov}, {Oberla}, {Oeyen}, {Pandya}, {Plaisier}, {Pyras}, {Ryckbosch}, {Scholten}, {Seckel}, {Smith}, {Southall}, {Torres}, {Toscano}, {Van Den Broeck}, {van Eijndhoven}, {Vieregg}, {Welling}, {Wissel}, {Young}, \& {Zink}}]{Aguilar+2021}
{Aguilar}, J.~A., {Allison}, P., {Beatty}, J.~J., {et~al.} 2021, Journal of Instrumentation, 16, P03025, \dodoi{10.1088/1748-0221/16/03/P03025}

\bibitem[{{\'A}lvarez-Mu{\~n}iz {et~al.}(2019){\'A}lvarez-Mu{\~n}iz, Alves~Batista, Balagopal~V., Bolmont, Bustamante, Carvalho, Charrier, Cognard, Decoene, Denton, De~Jong, De~Vries, Engel, Fang, Finley, Gabici, Gou, Gu, Gu{\'e}pin, Hu, Huang, Kotera, Le~Coz, Lenain, L{\"u}, Martineau-Huynh, Mostaf{\'a}, Mottez, Murase, Niess, Oikonomou, Pierog, Qian, Qin, Ran, Renault-Tinacci, Roth, Schr{\"o}der, Sch{\"u}ssler, Tasse, Timmermans, Tueros, Wu, Zarka, Zech, Zhang, Zhang, Zhang, Zheng, \& Zilles}]{GRANDO2020}
{\'A}lvarez-Mu{\~n}iz, J., Alves~Batista, R., Balagopal~V., A., {et~al.} 2019, Science China Physics, Mechanics \& Astronomy, 63, 219501, \dodoi{10.1007/s11433-018-9385-7}

\bibitem[{{Atwood} {et~al.}(2009){Atwood}, {Abdo}, {Ackermann}, {Althouse}, {Anderson}, {Axelsson}, {Baldini}, {Ballet}, {Band}, {Barbiellini}, {Bartelt}, {Bastieri}, {Baughman}, {Bechtol}, {B{\'e}d{\'e}r{\`e}de}, {Bellardi}, {Bellazzini}, {Berenji}, {Bignami}, {Bisello}, {Bissaldi}, {Blandford}, {Bloom}, {Bogart}, {Bonamente}, {Bonnell}, {Borgland}, {Bouvier}, {Bregeon}, {Brez}, {Brigida}, {Bruel}, {Burnett}, {Busetto}, {Caliandro}, {Cameron}, {Caraveo}, {Carius}, {Carlson}, {Casandjian}, {Cavazzuti}, {Ceccanti}, {Cecchi}, {Charles}, {Chekhtman}, {Cheung}, {Chiang}, {Chipaux}, {Cillis}, {Ciprini}, {Claus}, {Cohen-Tanugi}, {Condamoor}, {Conrad}, {Corbet}, {Corucci}, {Costamante}, {Cutini}, {Davis}, {Decotigny}, {DeKlotz}, {Dermer}, {de Angelis}, {Digel}, {do Couto e Silva}, {Drell}, {Dubois}, {Dumora}, {Edmonds}, {Fabiani}, {Farnier}, {Favuzzi}, {Flath}, {Fleury}, {Focke}, {Funk}, {Fusco}, {Gargano}, {Gasparrini}, {Gehrels}, {Gentit}, {Germani}, {Giebels}, {Giglietto}, {Giommi}, {Giordano}, {Glanzman},
  {Godfrey}, {Grenier}, {Grondin}, {Grove}, {Guillemot}, {Guiriec}, {Haller}, {Harding}, {Hart}, {Hays}, {Healey}, {Hirayama}, {Hjalmarsdotter}, {Horn}, {Hughes}, {J{\'o}hannesson}, {Johansson}, {Johnson}, {Johnson}, {Johnson}, {Johnson}, {Kamae}, {Katagiri}, {Kataoka}, {Kavelaars}, {Kawai}, {Kelly}, {Kerr}, {Klamra}, {Kn{\"o}dlseder}, {Kocian}, {Komin}, {Kuehn}, {Kuss}, {Landriu}, {Latronico}, {Lee}, {Lee}, {Lemoine-Goumard}, {Lionetto}, {Longo}, {Loparco}, {Lott}, {Lovellette}, {Lubrano}, {Madejski}, {Makeev}, {Marangelli}, {Massai}, {Mazziotta}, {McEnery}, {Menon}, {Meurer}, {Michelson}, {Minuti}, {Mirizzi}, {Mitthumsiri}, {Mizuno}, {Moiseev}, {Monte}, {Monzani}, {Moretti}, {Morselli}, {Moskalenko}, {Murgia}, {Nakamori}, {Nishino}, {Nolan}, {Norris}, {Nuss}, {Ohno}, {Ohsugi}, {Omodei}, {Orlando}, {Ormes}, {Paccagnella}, {Paneque}, {Panetta}, {Parent}, {Pearce}, {Pepe}, {Perazzo}, {Pesce-Rollins}, {Picozza}, {Pieri}, {Pinchera}, {Piron}, {Porter}, {Poupard}, {Rain{\`o}}, {Rando}, {Rapposelli}, {Razzano},
  {Reimer}, {Reimer}, {Reposeur}, {Reyes}, {Ritz}, {Rochester}, {Rodriguez}, {Romani}, {Roth}, {Russell}, {Ryde}, {Sabatini}, {Sadrozinski}, {Sanchez}, {Sander}, {Sapozhnikov}, {Parkinson}, {Scargle}, {Schalk}, {Scolieri}, {Sgr{\`o}}, {Share}, {Shaw}, {Shimokawabe}, {Shrader}, {Sierpowska-Bartosik}, {Siskind}, {Smith}, {Smith}, {Spandre}, {Spinelli}, {Starck}, {Stephens}, {Strickman}, {Strong}, {Suson}, {Tajima}, {Takahashi}, {Takahashi}, {Tanaka}, {Tenze}, {Tether}, {Thayer}, {Thayer}, {Thompson}, {Tibaldo}, {Tibolla}, {Torres}, {Tosti}, {Tramacere}, {Turri}, {Usher}, {Vilchez}, {Vitale}, {Wang}, {Watters}, {Winer}, {Wood}, {Ylinen}, \& {Ziegler}}]{Atwood2009}
{Atwood}, W.~B., {Abdo}, A.~A., {Ackermann}, M., {et~al.} 2009, \apj, 697, 1071, \dodoi{10.1088/0004-637X/697/2/1071}

\bibitem[{{Bisnovatyi-Kogan} \& {Ruzmaikin}(1974)}]{Bisnovatyi-Kogan1974}
{Bisnovatyi-Kogan}, G.~S., \& {Ruzmaikin}, A.~A. 1974, \apss, 28, 45, \dodoi{10.1007/BF00642237}

\bibitem[{{Blandford} {et~al.}(2019){Blandford}, {Meier}, \& {Readhead}}]{Blandford2019}
{Blandford}, R., {Meier}, D., \& {Readhead}, A. 2019, \araa, 57, 467, \dodoi{10.1146/annurev-astro-081817-051948}

\bibitem[{{Blandford} \& {Znajek}(1977)}]{Blandford1977}
{Blandford}, R.~D., \& {Znajek}, R.~L. 1977, \mnras, 179, 433, \dodoi{10.1093/mnras/179.3.433}

\bibitem[{Blumenthal \& Gould(1970)}]{Brumenthal1970}
Blumenthal, G.~R., \& Gould, R.~J. 1970, Rev. Mod. Phys., 42, 237, \dodoi{10.1103/RevModPhys.42.237}

\bibitem[{{Chen} {et~al.}(2023){Chen}, {Uzdensky}, \& {Dexter}}]{Chen+2022}
{Chen}, A.~Y., {Uzdensky}, D., \& {Dexter}, J. 2023, \apj, 944, 173, \dodoi{10.3847/1538-4357/acb68a}

\bibitem[{{Chen} {et~al.}(2018){Chen}, {Yuan}, \& {Yang}}]{Chen+2018}
{Chen}, A.~Y., {Yuan}, Y., \& {Yang}, H. 2018, \apjl, 863, L31, \dodoi{10.3847/2041-8213/aad8ab}

\bibitem[{{Chodorowski} {et~al.}(1992){Chodorowski}, {Zdziarski}, \& {Sikora}}]{Chodorowski1992}
{Chodorowski}, M.~J., {Zdziarski}, A.~A., \& {Sikora}, M. 1992, \apj, 400, 181, \dodoi{10.1086/171984}

\bibitem[{{Comisso} \& {Sironi}(2019)}]{ComissoSironi2019}
{Comisso}, L., \& {Sironi}, L. 2019, \apj, 886, 122, \dodoi{10.3847/1538-4357/ab4c33}

\bibitem[{{Comisso} {et~al.}(2020){Comisso}, {Sobacchi}, \& {Sironi}}]{Comisso+2020}
{Comisso}, L., {Sobacchi}, E., \& {Sironi}, L. 2020, in APS Meeting Abstracts, Vol. 2020, APS Division of Plasma Physics Meeting Abstracts, GO03.005

\bibitem[{{Crinquand} {et~al.}(2021){Crinquand}, {Cerutti}, {Dubus}, {Parfrey}, \& {Philippov}}]{Crinquand2021}
{Crinquand}, B., {Cerutti}, B., {Dubus}, G., {Parfrey}, K., \& {Philippov}, A. 2021, \aap, 650, A163, \dodoi{10.1051/0004-6361/202040158}

\bibitem[{{Davelaar} {et~al.}(2023){Davelaar}, {Ripperda}, {Sironi}, {Philippov}, {Olivares}, {Porth}, {Berg}, {Bronzwaer}, {Chatterjee}, \& {Liska}}]{Davelaar+2023}
{Davelaar}, J., {Ripperda}, B., {Sironi}, L., {et~al.} 2023, \apjl, 959, L3, \dodoi{10.3847/2041-8213/ad0b79}

\bibitem[{{Derby}(1978)}]{Derby1978}
{Derby}, N.~F., J. 1978, \apj, 224, 1013, \dodoi{10.1086/156451}

\bibitem[{{Dermer} \& {Menon}(2009)}]{Dermer2009}
{Dermer}, C.~D., \& {Menon}, G. 2009, {High Energy Radiation from Black Holes: Gamma Rays, Cosmic Rays, and Neutrinos} (PRINCETON UNIVERSITY PRESS)

\bibitem[{{EHT MWL Science Working Group} {et~al.}(2021){EHT MWL Science Working Group}, {Algaba}, {Anczarski}, {Asada}, {Balokovi{\'c}}, {Chandra}, {Cui}, {Falcone}, {Giroletti}, {Goddi}, {Hada}, {Haggard}, {Jorstad}, {Kaur}, {Kawashima}, {Keating}, {Kim}, {Kino}, {Komossa}, {Kravchenko}, {Krichbaum}, {Lee}, {Lu}, {Lucchini}, {Markoff}, {Neilsen}, {Nowak}, {Park}, {Principe}, {Ramakrishnan}, {Reynolds}, {Sasada}, {Savchenko}, {Williamson}, {Event Horizon Telescope Collaboration}, {Akiyama}, {Alberdi}, {Alef}, {Anantua}, {Azulay}, {Baczko}, {Ball}, {Barrett}, {Bintley}, {Benson}, {Blackburn}, {Blundell}, {Boland}, {Bouman}, {Bower}, {Boyce}, {Bremer}, {Brinkerink}, {Brissenden}, {Britzen}, {Broderick}, {Broguiere}, {Bronzwaer}, {Byun}, {Carlstrom}, {Chael}, {Chan}, {Chatterjee}, {Chatterjee}, {Chen}, {Chen}, {Chesler}, {Cho}, {Christian}, {Conway}, {Cordes}, {Crawford}, {Crew}, {Cruz-Osorio}, {Davelaar}, {de Laurentis}, {Deane}, {Dempsey}, {Desvignes}, {Dexter}, {Doeleman}, {Eatough}, {Falcke}, {Farah},
  {Fish}, {Fomalont}, {Ford}, {Fraga-Encinas}, {Friberg}, {Fromm}, {Fuentes}, {Galison}, {Gammie}, {Garc{\'\i}a}, {Gentaz}, {Georgiev}, {Gold}, {G{\'o}mez}, {G{\'o}mez-Ruiz}, {Gu}, {Gurwell}, {Hecht}, {Hesper}, {Ho}, {Ho}, {Honma}, {Huang}, {Huang}, {Hughes}, {Ikeda}, {Inoue}, {Issaoun}, {James}, {Jannuzi}, {Janssen}, {Jeter}, {Jiang}, {Jim{\'e}nez-Rosales}, {Johnson}, {Jung}, {Karami}, {Karuppusamy}, {Kettenis}, {Kim}, {Kim}, {Kim}, {Koay}, {Kofuji}, {Koch}, {Koyama}, {Kramer}, {Kramer}, {Kuo}, {Lauer}, {Levis}, {Li}, {Li}, {Lindqvist}, {Lico}, {Lindahl}, {Liu}, {Liu}, {Liuzzo}, {Lo}, {Lobanov}, {Loinard}, {Lonsdale}, {MacDonald}, {Mao}, {Marchili}, {Marrone}, {Marscher}, {Mart{\'\i}-Vidal}, {Matsushita}, {Matthews}, {Medeiros}, {Menten}, {Mizuno}, {Mizuno}, {Moran}, {Moriyama}, {Moscibrodzka}, {M{\"u}ller}, {Musoke}, {Mej{\'\i}as}, {Nagai}, {Nagar}, {Nakamura}, {Narayan}, {Narayanan}, {Natarajan}, {Nathanail}, {Neri}, {Ni}, {Noutsos}, {Okino}, {Olivares}, {Ortiz-Le{\'o}n}, {Oyama}, {{\"O}zel}, {Palumbo},
  {Patel}, {Pen}, {Pesce}, {Pi{\'e}tu}, {Plambeck}, {Popstefanija}, {Porth}, {P{\"o}tzl}, {Prather}, {Preciado-L{\'o}pez}, {Psaltis}, {Pu}, {Rao}, {Rawlings}, {Raymond}, {Rezzolla}, {Ricarte}, {Ripperda}, {Roelofs}, {Rogers}, {Ros}, {Rose}, {Roshanineshat}, {Rottmann}, {Roy}, {Ruszczyk}, {Rygl}, {S{\'a}nchez}, {S{\'a}nchez-Arguelles}, {Savolainen}, {Schloerb}, {Schuster}, {Shao}, {Shen}, {Small}, {Sohn}, {Soohoo}, {Sun}, {Tazaki}, {Tetarenko}, {Tiede}, {Tilanus}, {Titus}, {Toma}, {Torne}, {Trent}, {Traianou}, {Trippe}, {van Bemmel}, {van Langevelde}, {van Rossum}, {Wagner}, {Ward-Thompson}, {Wardle}, {Weintroub}, {Wex}, {Wharton}, {Wielgus}, {Wong}, {Wu}, {Yoon}, {Young}, {Young}, {Younsi}, {Yuan}, {Yuan}, {Zensus}, {Zhao}, {Zhao}, {Fermi Large Area Telescope Collaboration}, {Principe}, {Giroletti}, {D'Ammando}, {Orienti}, {H.~E.~S.~S. Collaboration}, {Abdalla}, {Adam}, {Aharonian}, {Benkhali}, {Ang{\"u}ner}, {Arcaro}, {Armand}, {Armstrong}, {Ashkar}, {Backes}, {Baghmanyan}, {Barbosa Martins}, {Barnacka},
  {Barnard}, {Becherini}, {Berge}, {Bernl{\"o}hr}, {Bi}, {B{\"o}ttcher}, {Boisson}, {Bolmont}, {de Lavergne}, {Breuhaus}, {Brun}, {Brun}, {Bryan}, {B{\"u}chele}, {Bulik}, {Bylund}, {Caroff}, {Carosi}, {Casanova}, {Chand}, {Chen}, {Cotter}, {Cury{\l}o}, {Damascene Mbarubucyeye}, {Davids}, {Davies}, {Deil}, {Devin}, {Dewilt}, {Dirson}, {Djannati-Ata{\"\i}}, {Dmytriiev}, {Donath}, {Doroshenko}, {Duffy}, {Dyks}, {Egberts}, {Eichhorn}, {Einecke}, {Emery}, {Ernenwein}, {Feijen}, {Fegan}, {Fiasson}, {de Clairfontaine}, {Fontaine}, {Funk}, {F{\"u}{\ss}ling}, {Gabici}, {Gallant}, {Giavitto}, {Giunti}, {Glawion}, {Glicenstein}, {Gottschall}, {Grondin}, {Hahn}, {Haupt}, {Hermann}, {Hinton}, {Hofmann}, {Hoischen}, {Holch}, {Holler}, {H{\"o}rbe}, {Horns}, {Huber}, {Jamrozy}, {Jankowsky}, {Jankowsky}, {Jardin-Blicq}, {Joshi}, {Jung-Richardt}, {Kasai}, {Kastendieck}, {Katarzy{\'n}ski}, {Katz}, {Khangulyan}, {Kh{\'e}lifi}, {Klepser}, {Klu{\'z}niak}, {Komin}, {Konno}, {Kosack}, {Kostunin}, {Kreter}, {Lamanna}, {Lemi{\`e}re},
  {Lemoine-Goumard}, {Lenain}, {Levy}, {Lohse}, {Lypova}, {Mackey}, {Majumdar}, {Malyshev}, {Malyshev}, {Marandon}, {Marchegiani}, {Marcowith}, {Mares}, {Mart{\'\i}-Devesa}, {Marx}, {Maurin}, {Meintjes}, {Meyer}, {Moderski}, {Mohamed}, {Mohrmann}, {Montanari}, {Moore}, {Morris}, {Moulin}, {Muller}, {Murach}, {Nakashima}, {Nayerhoda}, {de Naurois}, {Ndiyavala}, {Niederwanger}, {Niemiec}, {Oakes}, {O'Brien}, {Odaka}, {Ohm}, {Olivera-Nieto}, {de Ona Wilhelmi}, {Ostrowski}, {Panter}, {Panny}, {Parsons}, {Peron}, {Peyaud}, {Piel}, {Pita}, {Poireau}, {Noel}, {Prokhorov}, {Prokoph}, {P{\"u}hlhofer}, {Punch}, {Quirrenbach}, {Rauth}, {Reichherzer}, {Reimer}, {Reimer}, {Remy}, {Renaud}, {Rieger}, {Rinchiuso}, {Romoli}, {Rowell}, {Rudak}, {Ruiz-Velasco}, {Sahakian}, {Sailer}, {Sanchez}, {Santangelo}, {Sasaki}, {Scalici}, {Schutte}, {Schwanke}, {Schwemmer}, {Seglar-Arroyo}, {Senniappan}, {Seyffert}, {Shafi}, {Shiningayamwe}, {Simoni}, {Sinha}, {Sol}, {Specovius}, {Spencer}, {Spir-Jacob}, {Stawarz}, {Sun}, {Steenkamp},
  {Stegmann}, {Steinmassl}, {Steppa}, {Takahashi}, {Tavernier}, {Taylor}, {Terrier}, {Tiziani}, {Tluczykont}, {Tomankova}, {Trichard}, {Tsirou}, {Tuffs}, {Uchiyama}, {van der Walt}, {van Eldik}, {van Rensburg}, {van Soelen}, {Vasileiadis}, {Veh}, {Venter}, {Vincent}, {Vink}, {V{\"o}lk}, {Vuillaume}, {Wadiasingh}, {Wagner}, {Watson}, {Werner}, {White}, {Wierzcholska}, {Wong}, {Yusafzai}, {Zacharias}, {Zanin}, {Zargaryan}, {Zdziarski}, {Zech}, {Zhu}, {Zorn}, {Zouari}, {{\.Z}ywucka}, {MAGIC Collaboration}, {Acciari}, {Ansoldi}, {Antonelli}, {Engels}, {Artero}, {Asano}, {Baack}, {Babi{\'c}}, {Baquero}, {de Almeida}, {Barrio}, {Becerra Gonz{\'a}lez}, {Bednarek}, {Bellizzi}, {Bernardini}, {Bernardos}, {Berti}, {Besenrieder}, {Bhattacharyya}, {Bigongiari}, {Biland}, {Blanch}, {Bonnoli}, {Bo{\v{s}}njak}, {Busetto}, {Carosi}, {Ceribella}, {Cerruti}, {Chai}, {Chilingarian}, {Cikota}, {Colak}, {Colombo}, {Contreras}, {Cortina}, {Covino}, {D'Amico}, {D'Elia}, {da Vela}, {Dazzi}, {de Angelis}, {de Lotto}, {Delfino},
  {Delgado}, {Delgado Mendez}, {Depaoli}, {di Pierro}, {di Venere}, {Do Souto Espi{\~n}eira}, {Dominis Prester}, {Donini}, {Dorner}, {Doro}, {Elsaesser}, {Ramazani}, {Fattorini}, {Ferrara}, {Fonseca}, {Font}, {Fruck}, {Fukami}, {Garc{\'\i}a L{\'o}pez}, {Garczarczyk}, {Gasparyan}, {Gaug}, {Giglietto}, {Giordano}, {Gliwny}, {Godinovi{\'c}}, {Green}, {Green}, {Hadasch}, {Hahn}, {Heckmann}, {Herrera}, {Hoang}, {Hrupec}, {H{\"u}tten}, {Inada}, {Inoue}, {Ishio}, {Iwamura}, {Jim{\'e}nez}, {Jormanainen}, {Jouvin}, {Kajiwara}, {Karjalainen}, {Kerszberg}, {Kobayashi}, {Kubo}, {Kushida}, {Lamastra}, {Lelas}, {Leone}, {Lindfors}, {Lombardi}, {Longo}, {L{\'o}pez-Coto}, {L{\'o}pez-Moya}, {L{\'o}pez-Oramas}, {Loporchio}, {Machado de Oliveira Fraga}, {Maggio}, {Majumdar}, {Makariev}, {Mallamaci}, {Maneva}, {Manganaro}, {Mannheim}, {Maraschi}, {Mariotti}, {Mart{\'\i}nez}, {Mazin}, {Menchiari}, {Mender}, {Mi{\'c}anovi{\'c}}, {Miceli}, {Miener}, {Minev}, {Miranda}, {Mirzoyan}, {Molina}, {Moralejo}, {Morcuende}, {Moreno},
  {Moretti}, {Neustroev}, {Nigro}, {Nilsson}, {Nishijima}, {Noda}, {Nozaki}, {Ohtani}, {Oka}, {Otero-Santos}, {Paiano}, {Palatiello}, {Paneque}, {Paoletti}, {Paredes}, {Pavleti{\'c}}, {Pe{\~n}il}, {Perennes}, {Persic}, {Moroni}, {Prandini}, {Priyadarshi}, {Puljak}, {Rhode}, {Rib{\'o}}, {Rico}, {Righi}, {Rugliancich}, {Saha}, {Sahakyan}, {Saito}, {Sakurai}, {Satalecka}, {Saturni}, {Schleicher}, {Schmidt}, {Schweizer}, {Sitarek}, {{\v{S}}nidari{\'c}}, {Sobczynska}, {Spolon}, {Stamerra}, {Strom}, {Strzys}, {Suda}, {Suri{\'c}}, {Takahashi}, {Tavecchio}, {Temnikov}, {Terzi{\'c}}, {Teshima}, {Tosti}, {Truzzi}, {Tutone}, {Ubach}, {van Scherpenberg}, {Vanzo}, {Vazquez Acosta}, {Ventura}, {Verguilov}, {Vigorito}, {Vitale}, {Vovk}, {Will}, {Wunderlich}, {Zari{\'c}}, {VERITAS Collaboration}, {Adams}, {Benbow}, {Brill}, {Capasso}, {Christiansen}, {Chromey}, {Daniel}, {Errando}, {Farrell}, {Feng}, {Finley}, {Fortson}, {Furniss}, {Gent}, {Giuri}, {Hassan}, {Hervet}, {Holder}, {Hughes}, {Humensky}, {Jin}, {Kaaret},
  {Kertzman}, {Kieda}, {Kumar}, {Lang}, {Lundy}, {Maier}, {Moriarty}, {Mukherjee}, {Nieto}, {Nievas-Rosillo}, {O'Brien}, {Ong}, {Otte}, {Patel}, {Pfrang}, {Pohl}, {Prado}, {Pueschel}, {Quinn}, {Ragan}, {Reynolds}, {Ribeiro}, {Richards}, {Roache}, {Rulten}, {Ryan}, {Santander}, {Sembroski}, {Shang}, {Weinstein}, {Williams}, {Williamson}, {Eavn Collaboration}, {Hirota}, {Cui}, {Niinuma}, {Ro}, {Sakai}, {Sawada-Satoh}, {Wajima}, {Wang}, {Liu}, \& {Yonekura}}]{EHT_MWL2021}
{EHT MWL Science Working Group}, {Algaba}, J.~C., {Anczarski}, J., {et~al.} 2021, \apjl, 911, L11, \dodoi{10.3847/2041-8213/abef71}

\bibitem[{{Event Horizon Telescope Collaboration} {et~al.}(2021{\natexlab{a}}){Event Horizon Telescope Collaboration}, {Akiyama}, {Algaba}, {Alberdi}, {Alef}, {Anantua}, {Asada}, {Azulay}, {Baczko}, {Ball}, {Balokovi{\'c}}, {Barrett}, {Benson}, {Bintley}, {Blackburn}, {Blundell}, {Boland}, {Bouman}, {Bower}, {Boyce}, {Bremer}, {Brinkerink}, {Brissenden}, {Britzen}, {Broderick}, {Broguiere}, {Bronzwaer}, {Byun}, {Carlstrom}, {Chael}, {Chan}, {Chatterjee}, {Chatterjee}, {Chen}, {Chen}, {Chesler}, {Cho}, {Christian}, {Conway}, {Cordes}, {Crawford}, {Crew}, {Cruz-Osorio}, {Cui}, {Davelaar}, {De Laurentis}, {Deane}, {Dempsey}, {Desvignes}, {Dexter}, {Doeleman}, {Eatough}, {Falcke}, {Farah}, {Fish}, {Fomalont}, {Ford}, {Fraga-Encinas}, {Freeman}, {Friberg}, {Fromm}, {Fuentes}, {Galison}, {Gammie}, {Garc{\'\i}a}, {Gentaz}, {Georgiev}, {Goddi}, {Gold}, {G{\'o}mez}, {G{\'o}mez-Ruiz}, {Gu}, {Gurwell}, {Hada}, {Haggard}, {Hecht}, {Hesper}, {Ho}, {Ho}, {Honma}, {Huang}, {Huang}, {Hughes}, {Ikeda}, {Inoue}, {Issaoun},
  {James}, {Jannuzi}, {Janssen}, {Jeter}, {Jiang}, {Jimenez-Rosales}, {Johnson}, {Jorstad}, {Jung}, {Karami}, {Karuppusamy}, {Kawashima}, {Keating}, {Kettenis}, {Kim}, {Kim}, {Kim}, {Kim}, {Kino}, {Koay}, {Kofuji}, {Koch}, {Koyama}, {Kramer}, {Kramer}, {Krichbaum}, {Kuo}, {Lauer}, {Lee}, {Levis}, {Li}, {Li}, {Lindqvist}, {Lico}, {Lindahl}, {Liu}, {Liu}, {Liuzzo}, {Lo}, {Lobanov}, {Loinard}, {Lonsdale}, {Lu}, {MacDonald}, {Mao}, {Marchili}, {Markoff}, {Marrone}, {Marscher}, {Mart{\'\i}-Vidal}, {Matsushita}, {Matthews}, {Medeiros}, {Menten}, {Mizuno}, {Mizuno}, {Moran}, {Moriyama}, {Moscibrodzka}, {M{\"u}ller}, {Musoke}, {Mej{\'\i}as}, {Michalik}, {Nadolski}, {Nagai}, {Nagar}, {Nakamura}, {Narayan}, {Narayanan}, {Natarajan}, {Nathanail}, {Neilsen}, {Neri}, {Ni}, {Noutsos}, {Nowak}, {Okino}, {Olivares}, {Ortiz-Le{\'o}n}, {Oyama}, {{\"O}zel}, {Palumbo}, {Park}, {Patel}, {Pen}, {Pesce}, {Pi{\'e}tu}, {Plambeck}, {PopStefanija}, {Porth}, {P{\"o}tzl}, {Prather}, {Preciado-L{\'o}pez}, {Psaltis}, {Pu}, {Ramakrishnan},
  {Rao}, {Rawlings}, {Raymond}, {Rezzolla}, {Ricarte}, {Ripperda}, {Roelofs}, {Rogers}, {Ros}, {Rose}, {Roshanineshat}, {Rottmann}, {Roy}, {Ruszczyk}, {Rygl}, {S{\'a}nchez}, {S{\'a}nchez-Arguelles}, {Sasada}, {Savolainen}, {Schloerb}, {Schuster}, {Shao}, {Shen}, {Small}, {Sohn}, {SooHoo}, {Sun}, {Tazaki}, {Tetarenko}, {Tiede}, {Tilanus}, {Titus}, {Toma}, {Torne}, {Trent}, {Traianou}, {Trippe}, {van Bemmel}, {van Langevelde}, {van Rossum}, {Wagner}, {Ward-Thompson}, {Wardle}, {Weintroub}, {Wex}, {Wharton}, {Wielgus}, {Wong}, {Wu}, {Yoon}, {Young}, {Young}, {Younsi}, {Yuan}, {Yuan}, {Zensus}, {Zhao}, \& {Zhao}}]{EHT2021VII}
{Event Horizon Telescope Collaboration}, {Akiyama}, K., {Algaba}, J.~C., {et~al.} 2021{\natexlab{a}}, \apjl, 910, L12, \dodoi{10.3847/2041-8213/abe71d}

\bibitem[{{Event Horizon Telescope Collaboration} {et~al.}(2021{\natexlab{b}}){Event Horizon Telescope Collaboration}, {Akiyama}, {Algaba}, {Alberdi}, {Alef}, {Anantua}, {Asada}, {Azulay}, {Baczko}, {Ball}, {Balokovi{\'c}}, {Barrett}, {Benson}, {Bintley}, {Blackburn}, {Blundell}, {Boland}, {Bouman}, {Bower}, {Boyce}, {Bremer}, {Brinkerink}, {Brissenden}, {Britzen}, {Broderick}, {Broguiere}, {Bronzwaer}, {Byun}, {Carlstrom}, {Chael}, {Chan}, {Chatterjee}, {Chatterjee}, {Chen}, {Chen}, {Chesler}, {Cho}, {Christian}, {Conway}, {Cordes}, {Crawford}, {Crew}, {Cruz-Osorio}, {Cui}, {Davelaar}, {De Laurentis}, {Deane}, {Dempsey}, {Desvignes}, {Dexter}, {Doeleman}, {Eatough}, {Falcke}, {Farah}, {Fish}, {Fomalont}, {Ford}, {Fraga-Encinas}, {Friberg}, {Fromm}, {Fuentes}, {Galison}, {Gammie}, {Garc{\'\i}a}, {Gelles}, {Gentaz}, {Georgiev}, {Goddi}, {Gold}, {G{\'o}mez}, {G{\'o}mez-Ruiz}, {Gu}, {Gurwell}, {Hada}, {Haggard}, {Hecht}, {Hesper}, {Himwich}, {Ho}, {Ho}, {Honma}, {Huang}, {Huang}, {Hughes}, {Ikeda}, {Inoue},
  {Issaoun}, {James}, {Jannuzi}, {Janssen}, {Jeter}, {Jiang}, {Jimenez-Rosales}, {Johnson}, {Jorstad}, {Jung}, {Karami}, {Karuppusamy}, {Kawashima}, {Keating}, {Kettenis}, {Kim}, {Kim}, {Kim}, {Kim}, {Kino}, {Koay}, {Kofuji}, {Koch}, {Koyama}, {Kramer}, {Kramer}, {Krichbaum}, {Kuo}, {Lauer}, {Lee}, {Levis}, {Li}, {Li}, {Lindqvist}, {Lico}, {Lindahl}, {Liu}, {Liu}, {Liuzzo}, {Lo}, {Lobanov}, {Loinard}, {Lonsdale}, {Lu}, {MacDonald}, {Mao}, {Marchili}, {Markoff}, {Marrone}, {Marscher}, {Mart{\'\i}-Vidal}, {Matsushita}, {Matthews}, {Medeiros}, {Menten}, {Mizuno}, {Mizuno}, {Moran}, {Moriyama}, {Moscibrodzka}, {M{\"u}ller}, {Musoke}, {Mus Mej{\'\i}as}, {Michalik}, {Nadolski}, {Nagai}, {Nagar}, {Nakamura}, {Narayan}, {Narayanan}, {Natarajan}, {Nathanail}, {Neilsen}, {Neri}, {Ni}, {Noutsos}, {Nowak}, {Okino}, {Olivares}, {Ortiz-Le{\'o}n}, {Oyama}, {{\"O}zel}, {Palumbo}, {Park}, {Patel}, {Pen}, {Pesce}, {Pi{\'e}tu}, {Plambeck}, {PopStefanija}, {Porth}, {P{\"o}tzl}, {Prather}, {Preciado-L{\'o}pez}, {Psaltis}, {Pu},
  {Ramakrishnan}, {Rao}, {Rawlings}, {Raymond}, {Rezzolla}, {Ricarte}, {Ripperda}, {Roelofs}, {Rogers}, {Ros}, {Rose}, {Roshanineshat}, {Rottmann}, {Roy}, {Ruszczyk}, {Rygl}, {S{\'a}nchez}, {S{\'a}nchez-Arguelles}, {Sasada}, {Savolainen}, {Schloerb}, {Schuster}, {Shao}, {Shen}, {Small}, {Sohn}, {SooHoo}, {Sun}, {Tazaki}, {Tetarenko}, {Tiede}, {Tilanus}, {Titus}, {Toma}, {Torne}, {Trent}, {Traianou}, {Trippe}, {van Bemmel}, {van Langevelde}, {van Rossum}, {Wagner}, {Ward-Thompson}, {Wardle}, {Weintroub}, {Wex}, {Wharton}, {Wielgus}, {Wong}, {Wu}, {Yoon}, {Young}, {Young}, {Younsi}, {Yuan}, {Yuan}, {Zensus}, {Zhao}, \& {Zhao}}]{EHT2021VIII}
---. 2021{\natexlab{b}}, \apjl, 910, L13, \dodoi{10.3847/2041-8213/abe4de}

\bibitem[{{Galeev} \& {Oraevskii}(1963)}]{GaleevOraevskii1963}
{Galeev}, A.~A., \& {Oraevskii}, V.~N. 1963, Soviet Physics Doklady, 7, 988

\bibitem[{{Goto} \& {Asano}(2022)}]{GotoAsano2022}
{Goto}, R., \& {Asano}, K. 2022, \apj, 933, 18, \dodoi{10.3847/1538-4357/ac67d5}

\bibitem[{Guo {et~al.}(2014)Guo, Li, Daughton, \& Liu}]{Guo+2014}
Guo, F., Li, H., Daughton, W., \& Liu, Y.-H. 2014, Phys. Rev. Lett., 113, 155005, \dodoi{10.1103/PhysRevLett.113.155005}

\bibitem[{{Guo} {et~al.}(2015){Guo}, {Liu}, {Daughton}, \& {Li}}]{Guo+2015}
{Guo}, F., {Liu}, Y.-H., {Daughton}, W., \& {Li}, H. 2015, \apj, 806, 167, \dodoi{10.1088/0004-637X/806/2/167}

\bibitem[{{Guo} {et~al.}(2020){Guo}, {Liu}, {Li}, {Li}, {Daughton}, \& {Kilian}}]{Guo2020}
{Guo}, F., {Liu}, Y.-H., {Li}, X., {et~al.} 2020, Physics of Plasmas, 27, 080501, \dodoi{10.1063/5.0012094}

\bibitem[{{Guo} {et~al.}(2016){Guo}, {Li}, {Li}, {Daughton}, {Zhang}, {Lloyd-Ronning}, {Liu}, {Zhang}, \& {Deng}}]{Guo2016}
{Guo}, F., {Li}, X., {Li}, H., {et~al.} 2016, \apjl, 818, L9, \dodoi{10.3847/2041-8205/818/1/L9}

\bibitem[{{Hada}(2019)}]{Hada2019}
{Hada}, K. 2019, Galaxies, 8, 1, \dodoi{10.3390/galaxies8010001}

\bibitem[{{Hada} {et~al.}(2013){Hada}, {Kino}, {Doi}, {Nagai}, {Honma}, {Hagiwara}, {Giroletti}, {Giovannini}, \& {Kawaguchi}}]{Hada2013}
{Hada}, K., {Kino}, M., {Doi}, A., {et~al.} 2013, \apj, 775, 70, \dodoi{10.1088/0004-637X/775/1/70}

\bibitem[{{Hakobyan} {et~al.}(2021){Hakobyan}, {Petropoulou}, {Spitkovsky}, \& {Sironi}}]{Hakobyan2021}
{Hakobyan}, H., {Petropoulou}, M., {Spitkovsky}, A., \& {Sironi}, L. 2021, \apj, 912, 48, \dodoi{10.3847/1538-4357/abedac}

\bibitem[{{Hakobyan} {et~al.}(2023){Hakobyan}, {Ripperda}, \& {Philippov}}]{Hakobyan2023}
{Hakobyan}, H., {Ripperda}, B., \& {Philippov}, A.~A. 2023, \apjl, 943, L29, \dodoi{10.3847/2041-8213/acb264}

\bibitem[{Hallmann {et~al.}(2021)Hallmann, Clark, Glaser, \& Smith}]{Gen2-Radio2021}
Hallmann, S., Clark, B., Glaser, C., \& Smith, D. 2021, Sensitivity studies for the IceCube-Gen2 radio array.
\newblock \doarXiv{2107.08910}

\bibitem[{{Hoshino}(2012)}]{HoshinoPhyLV2012}
{Hoshino}, M. 2012, \prl, 108, 135003, \dodoi{10.1103/PhysRevLett.108.135003}

\bibitem[{{Hoshino} \& {Lyubarsky}(2012)}]{Hoshino2012}
{Hoshino}, M., \& {Lyubarsky}, Y. 2012, \ssr, 173, 521, \dodoi{10.1007/s11214-012-9931-z}

\bibitem[{{Imazawa} {et~al.}(2021){Imazawa}, {Fukazawa}, \& {Takahashi}}]{Imazawa2021}
{Imazawa}, R., {Fukazawa}, Y., \& {Takahashi}, H. 2021, \apj, 919, 110, \dodoi{10.3847/1538-4357/ac0ae4}

\bibitem[{{Ishizaki} \& {Ioka}(2024)}]{Ishizaki2024}
{Ishizaki}, W., \& {Ioka}, K. 2024, \pre, 110, 015205, \dodoi{10.1103/PhysRevE.110.015205}

\bibitem[{{Jiang} {et~al.}(2021){Jiang}, {Shen}, {Mart{\'\i}-Vidal}, {Wang}, {Jiang}, \& {Kawaguchi}}]{Jiang2021}
{Jiang}, W., {Shen}, Z., {Mart{\'\i}-Vidal}, I., {et~al.} 2021, \apjl, 922, L16, \dodoi{10.3847/2041-8213/ac375c}

\bibitem[{{Kawazura} {et~al.}(2020){Kawazura}, {Schekochihin}, {Barnes}, {TenBarge}, {Tong}, {Klein}, \& {Dorland}}]{Kawazura2020}
{Kawazura}, Y., {Schekochihin}, A.~A., {Barnes}, M., {et~al.} 2020, Physical Review X, 10, 041050, \dodoi{10.1103/PhysRevX.10.041050}

\bibitem[{{Kimura} {et~al.}(2019{\natexlab{a}}){Kimura}, {Murase}, \& {M{\'e}sz{\'a}ros}}]{KimuraMurase2019}
{Kimura}, S.~S., {Murase}, K., \& {M{\'e}sz{\'a}ros}, P. 2019{\natexlab{a}}, \prd, 100, 083014, \dodoi{10.1103/PhysRevD.100.083014}

\bibitem[{{Kimura} {et~al.}(2015){Kimura}, {Murase}, \& {Toma}}]{Kimura2015}
{Kimura}, S.~S., {Murase}, K., \& {Toma}, K. 2015, \apj, 806, 159, \dodoi{10.1088/0004-637X/806/2/159}

\bibitem[{{Kimura} \& {Toma}(2020)}]{Kimura2020}
{Kimura}, S.~S., \& {Toma}, K. 2020, \apj, 905, 178, \dodoi{10.3847/1538-4357/abc343}

\bibitem[{{Kimura} {et~al.}(2022){Kimura}, {Toma}, {Noda}, \& {Hada}}]{Kimura+2022}
{Kimura}, S.~S., {Toma}, K., {Noda}, H., \& {Hada}, K. 2022, \apjl, 937, L34, \dodoi{10.3847/2041-8213/ac8d5a}

\bibitem[{{Kimura} {et~al.}(2016){Kimura}, {Toma}, {Suzuki}, \& {Inutsuka}}]{Kimura2016}
{Kimura}, S.~S., {Toma}, K., {Suzuki}, T.~K., \& {Inutsuka}, S.-i. 2016, \apj, 822, 88, \dodoi{10.3847/0004-637X/822/2/88}

\bibitem[{{Kimura} {et~al.}(2019{\natexlab{b}}){Kimura}, {Tomida}, \& {Murase}}]{KimuraTomida2019}
{Kimura}, S.~S., {Tomida}, K., \& {Murase}, K. 2019{\natexlab{b}}, \mnras, 485, 163, \dodoi{10.1093/mnras/stz329}

\bibitem[{{Kin} {et~al.}(2024){Kin}, {Kisaka}, {Toma}, {Kimura}, \& {Levinson}}]{Kin2024}
{Kin}, K., {Kisaka}, S., {Toma}, K., {Kimura}, S.~S., \& {Levinson}, A. 2024, \apj, 964, 78, \dodoi{10.3847/1538-4357/ad20cd}

\bibitem[{{Kino} {et~al.}(2015){Kino}, {Takahara}, {Hada}, {Akiyama}, {Nagai}, \& {Sohn}}]{Kino2015}
{Kino}, M., {Takahara}, F., {Hada}, K., {et~al.} 2015, \apj, 803, 30, \dodoi{10.1088/0004-637X/803/1/30}

\bibitem[{{Kisaka} {et~al.}(2020){Kisaka}, {Levinson}, \& {Toma}}]{Kisaka2020}
{Kisaka}, S., {Levinson}, A., \& {Toma}, K. 2020, \apj, 902, 80, \dodoi{10.3847/1538-4357/abb46c}

\bibitem[{{Kulsrud}(2005)}]{Kulsrud2005}
{Kulsrud}, R.~M. 2005, {Plasma Physics for Astrophysics}

\bibitem[{{Kuze} {et~al.}(2022){Kuze}, {Kimura}, \& {Toma}}]{Kuze+2022}
{Kuze}, R., {Kimura}, S.~S., \& {Toma}, K. 2022, \apj, 935, 159, \dodoi{10.3847/1538-4357/ac7ec1}

\bibitem[{{Levinson} \& {Cerutti}(2018)}]{LevinsonCerutti2018}
{Levinson}, A., \& {Cerutti}, B. 2018, \aap, 616, A184, \dodoi{10.1051/0004-6361/201832915}

\bibitem[{{Lucchini} {et~al.}(2019){Lucchini}, {Krau{\ss}}, \& {Markoff}}]{Lucchini+2019}
{Lucchini}, M., {Krau{\ss}}, F., \& {Markoff}, S. 2019, \mnras, 489, 1633, \dodoi{10.1093/mnras/stz2125}

\bibitem[{{Lynn} {et~al.}(2014){Lynn}, {Quataert}, {Chandran}, \& {Parrish}}]{Lynn+2014}
{Lynn}, J.~W., {Quataert}, E., {Chandran}, B. D.~G., \& {Parrish}, I.~J. 2014, \apj, 791, 71, \dodoi{10.1088/0004-637X/791/1/71}

\bibitem[{{MAGIC Collaboration} {et~al.}(2020){MAGIC Collaboration}, {Acciari}, {Ansoldi}, {Antonelli}, {Arbet Engels}, {Arcaro}, {Baack}, {Babi{\'c}}, {Banerjee}, {Bangale}, {Barres de Almeida}, {Barrio}, {Becerra Gonz{\'a}lez}, {Bednarek}, {Bellizzi}, {Bernardini}, {Berti}, {Besenrieder}, {Bhattacharyya}, {Bigongiari}, {Biland}, {Blanch}, {Bonnoli}, {Bo{\v{s}}njak}, {Busetto}, {Carosi}, {Ceribella}, {Chai}, {Chilingaryan}, {Cikota}, {Colak}, {Colin}, {Colombo}, {Contreras}, {Cortina}, {Covino}, {D'Elia}, {da Vela}, {Dazzi}, {de Angelis}, {de Lotto}, {Delfino}, {Delgado}, {Depaoli}, {di Pierro}, {di Venere}, {Do Souto Espi{\~n}eira}, {Dominis Prester}, {Donini}, {Dorner}, {Doro}, {Elsaesser}, {Fallah Ramazani}, {Fattorini}, {Fern{\'a}ndez-Barral}, {Ferrara}, {Fidalgo}, {Foffano}, {Fonseca}, {Font}, {Fruck}, {Fukami}, {Garc{\'\i}a L{\'o}pez}, {Garczarczyk}, {Gasparyan}, {Gaug}, {Giglietto}, {Giordano}, {Godinovi{\'c}}, {Green}, {Guberman}, {Hadasch}, {Hahn}, {Herrera}, {Hoang}, {Hrupec}, {H{\"u}tten},
  {Inada}, {Inoue}, {Ishio}, {Iwamura}, {Jouvin}, {Kerszberg}, {Kubo}, {Kushida}, {Lamastra}, {Lelas}, {Leone}, {Lindfors}, {Lombardi}, {Longo}, {L{\'o}pez}, {L{\'o}pez-Coto}, {L{\'o}pez-Oramas}, {Loporchio}, {Machado de Oliveira Fraga}, {Maggio}, {Majumdar}, {Makariev}, {Mallamaci}, {Maneva}, {Manganaro}, {Mannheim}, {Maraschi}, {Mariotti}, {Mart{\'\i}nez}, {Masuda}, {Mazin}, {Mi{\'c}anovi{\'c}}, {Miceli}, {Minev}, {Miranda}, {Mirzoyan}, {Molina}, {Moralejo}, {Morcuende}, {Moreno}, {Moretti}, {Munar-Adrover}, {Neustroev}, {Nigro}, {Nilsson}, {Ninci}, {Nishijima}, {Noda}, {Nogu{\'e}s}, {N{\"o}the}, {Nozaki}, {Paiano}, {Palacio}, {Palatiello}, {Paneque}, {Paoletti}, {Paredes}, {Pe{\~n}il}, {Peresano}, {Persic}, {Prada Moroni}, {Prandini}, {Puljak}, {Rhode}, {Rib{\'o}}, {Rico}, {Righi}, {Rugliancich}, {Saha}, {Sahakyan}, {Saito}, {Sakurai}, {Satalecka}, {Schmidt}, {Schweizer}, {Sitarek}, {{\v{S}}nidari{\'c}}, {Sobczynska}, {Somero}, {Stamerra}, {Strom}, {Strzys}, {Suda}, {Suri{\'c}}, {Takahashi}, {Tavecchio},
  {Temnikov}, {Terzi{\'c}}, {Teshima}, {Torres-Alb{\`a}}, {Tosti}, {Tsujimoto}, {Vagelli}, {van Scherpenberg}, {Vanzo}, {Acosta}, {Vigorito}, {Vitale}, {Vovk}, {Will}, {Zari{\'c}}, {Asano}, {Hada}, {Harris}, {Giroletti}, {Jermak}, {Madrid}, {Massaro}, {Richter}, {Spanier}, {Steele}, \& {Walker}}]{MAGIC2020}
{MAGIC Collaboration}, {Acciari}, V.~A., {Ansoldi}, S., {et~al.} 2020, \mnras, 492, 5354, \dodoi{10.1093/mnras/staa014}

\bibitem[{{Marshall} {et~al.}(2018){Marshall}, {Avara}, \& {McKinney}}]{Marshall+2018}
{Marshall}, M.~D., {Avara}, M.~J., \& {McKinney}, J.~C. 2018, \mnras, 478, 1837, \dodoi{10.1093/mnras/sty1184}

\bibitem[{{McKinney} {et~al.}(2012){McKinney}, {Tchekhovskoy}, \& {Blandford}}]{Mckinney2012}
{McKinney}, J.~C., {Tchekhovskoy}, A., \& {Blandford}, R.~D. 2012, \mnras, 423, 3083, \dodoi{10.1111/j.1365-2966.2012.21074.x}

\bibitem[{{Mehlhaff} {et~al.}(2024){Mehlhaff}, {Werner}, {Cerutti}, {Uzdensky}, \& {Begelman}}]{Mehlhaff2024}
{Mehlhaff}, J., {Werner}, G., {Cerutti}, B., {Uzdensky}, D., \& {Begelman}, M. 2024, \mnras, 527, 11587, \dodoi{10.1093/mnras/stad3863}

\bibitem[{{Mo{\'s}cibrodzka} {et~al.}(2016){Mo{\'s}cibrodzka}, {Falcke}, \& {Shiokawa}}]{Moscibrodzka2016}
{Mo{\'s}cibrodzka}, M., {Falcke}, H., \& {Shiokawa}, H. 2016, \aap, 586, A38, \dodoi{10.1051/0004-6361/201526630}

\bibitem[{{Murase} \& {Nagataki}(2006)}]{MuraseNagataki2006b}
{Murase}, K., \& {Nagataki}, S. 2006, \prd, 73, 063002, \dodoi{10.1103/PhysRevD.73.063002}

\bibitem[{{Nakamura} {et~al.}(2018){Nakamura}, {Asada}, {Hada}, {Pu}, {Noble}, {Tseng}, {Toma}, {Kino}, {Nagai}, {Takahashi}, {Algaba}, {Orienti}, {Akiyama}, {Doi}, {Giovannini}, {Giroletti}, {Honma}, {Koyama}, {Lico}, {Niinuma}, \& {Tazaki}}]{Nakamura2018}
{Nakamura}, M., {Asada}, K., {Hada}, K., {et~al.} 2018, \apj, 868, 146, \dodoi{10.3847/1538-4357/aaeb2d}

\bibitem[{{Narayan} {et~al.}(2012){Narayan}, {S{\"A} dowski}, {Penna}, \& {Kulkarni}}]{Narayan2012}
{Narayan}, R., {S{\"A} dowski}, A., {Penna}, R.~F., \& {Kulkarni}, A.~K. 2012, \mnras, 426, 3241, \dodoi{10.1111/j.1365-2966.2012.22002.x}

\bibitem[{{Narayan} \& {Yi}(1994)}]{Narayah&Yi1994}
{Narayan}, R., \& {Yi}, I. 1994, \apjl, 428, L13, \dodoi{10.1086/187381}

\bibitem[{{N{\"a}ttil{\"a}} \& {Beloborodov}(2022)}]{Nattila2022}
{N{\"a}ttil{\"a}}, J., \& {Beloborodov}, A.~M. 2022, \prl, 128, 075101, \dodoi{10.1103/PhysRevLett.128.075101}

\bibitem[{{Park} {et~al.}(2019){Park}, {Hada}, {Kino}, {Nakamura}, {Hodgson}, {Ro}, {Cui}, {Asada}, {Algaba}, {Sawada-Satoh}, {Lee}, {Cho}, {Shen}, {Jiang}, {Trippe}, {Niinuma}, {Sohn}, {Jung}, {Zhao}, {Wajima}, {Tazaki}, {Honma}, {An}, {Akiyama}, {Byun}, {Kim}, {Zhang}, {Cheng}, {Kobayashi}, {Shibata}, {Lee}, {Roh}, {Oh}, {Yeom}, {Jung}, {Oh}, {Kim}, {Hwang}, \& {Hagiwara}}]{Park2019}
{Park}, J., {Hada}, K., {Kino}, M., {et~al.} 2019, \apj, 887, 147, \dodoi{10.3847/1538-4357/ab5584}

\bibitem[{{Petropoulou} {et~al.}(2019){Petropoulou}, {Sironi}, {Spitkovsky}, \& {Giannios}}]{Petropoulou+2019}
{Petropoulou}, M., {Sironi}, L., {Spitkovsky}, A., \& {Giannios}, D. 2019, \apj, 880, 37, \dodoi{10.3847/1538-4357/ab287a}

\bibitem[{{Porth} {et~al.}(2019){Porth}, {Chatterjee}, {Narayan}, {Gammie}, {Mizuno}, {Anninos}, {Baker}, {Bugli}, {Chan}, {Davelaar}, {Del Zanna}, {Etienne}, {Fragile}, {Kelly}, {Liska}, {Markoff}, {McKinney}, {Mishra}, {Noble}, {Olivares}, {Prather}, {Rezzolla}, {Ryan}, {Stone}, {Tomei}, {White}, {Younsi}, {Akiyama}, {Alberdi}, {Alef}, {Asada}, {Azulay}, {Baczko}, {Ball}, {Balokovi{\'c}}, {Barrett}, {Bintley}, {Blackburn}, {Boland}, {Bouman}, {Bower}, {Bremer}, {Brinkerink}, {Brissenden}, {Britzen}, {Broderick}, {Broguiere}, {Bronzwaer}, {Byun}, {Carlstrom}, {Chael}, {Chatterjee}, {Chen}, {Chen}, {Cho}, {Christian}, {Conway}, {Cordes}, {Geoffrey}, {Crew}, {Cui}, {De Laurentis}, {Deane}, {Dempsey}, {Desvignes}, {Doeleman}, {Eatough}, {Falcke}, {Fish}, {Fomalont}, {Fraga-Encinas}, {Freeman}, {Friberg}, {Fromm}, {G{\'o}mez}, {Galison}, {Garc{\'\i}a}, {Gentaz}, {Georgiev}, {Goddi}, {Gold}, {Gu}, {Gurwell}, {Hada}, {Hecht}, {Hesper}, {Ho}, {Ho}, {Honma}, {Huang}, {Huang}, {Hughes}, {Ikeda}, {Inoue}, {Issaoun},
  {James}, {Jannuzi}, {Janssen}, {Jeter}, {Jiang}, {Johnson}, {Jorstad}, {Jung}, {Karami}, {Karuppusamy}, {Kawashima}, {Keating}, {Kettenis}, {Kim}, {Kim}, {Kim}, {Kino}, {Koay}, {Patrick}, {Koch}, {Koyama}, {Kramer}, {Kramer}, {Krichbaum}, {Kuo}, {Lauer}, {Lee}, {Li}, {Li}, {Lindqvist}, {Liu}, {Liuzzo}, {Lo}, {Lobanov}, {Loinard}, {Lonsdale}, {Lu}, {MacDonald}, {Mao}, {Marrone}, {Marscher}, {Mart{\'\i}-Vidal}, {Matsushita}, {Matthews}, {Medeiros}, {Menten}, {Mizuno}, {Moran}, {Moriyama}, {Moscibrodzka}, {M{\"u}ller}, {Nagai}, {Nagar}, {Nakamura}, {Narayanan}, {Natarajan}, {Neri}, {Ni}, {Noutsos}, {Okino}, {Oyama}, {{\"O}zel}, {Palumbo}, {Patel}, {Pen}, {Pesce}, {Pi{\'e}tu}, {Plambeck}, {PopStefanija}, {Preciado-L{\'o}pez}, {Psaltis}, {Pu}, {Ramakrishnan}, {Rao}, {Rawlings}, {Raymond}, {Ripperda}, {Roelofs}, {Rogers}, {Ros}, {Rose}, {Roshanineshat}, {Rottmann}, {Roy}, {Ruszczyk}, {Rygl}, {S{\'a}nchez}, {S{\'a}nchez-Arguelles}, {Sasada}, {Savolainen}, {Schloerb}, {Schuster}, {Shao}, {Shen}, {Small}, {Sohn},
  {SooHoo}, {Tazaki}, {Tiede}, {Tilanus}, {Titus}, {Toma}, {Torne}, {Trent}, {Trippe}, {Tsuda}, {van Bemmel}, {van Langevelde}, {van Rossum}, {Wagner}, {Wardle}, {Weintroub}, {Wex}, {Wharton}, {Wielgus}, {Wong}, {Wu}, {Young}, {Young}, {Yuan}, {Yuan}, {Zensus}, {Zhao}, {Zhao}, {Zhu}, \& {Event Horizon Telescope Collaboration}}]{Porth+2019}
{Porth}, O., {Chatterjee}, K., {Narayan}, R., {et~al.} 2019, \apjs, 243, 26, \dodoi{10.3847/1538-4365/ab29fd}

\bibitem[{{Ripperda} {et~al.}(2022){Ripperda}, {Liska}, {Chatterjee}, {Musoke}, {Philippov}, {Markoff}, {Tchekhovskoy}, \& {Younsi}}]{Ripperda2022}
{Ripperda}, B., {Liska}, M., {Chatterjee}, K., {et~al.} 2022, \apjl, 924, L32, \dodoi{10.3847/2041-8213/ac46a1}

\bibitem[{{Sagdeev} \& {Galeev}(1969)}]{SagdeevGaleev1969}
{Sagdeev}, R.~Z., \& {Galeev}, A.~A. 1969, {Nonlinear Plasma Theory}

\bibitem[{{San} \& {Maulik}(2018)}]{SanMaulik2018}
{San}, O., \& {Maulik}, R. 2018, Nonlinear Processes in Geophysics, 25, 457, \dodoi{10.5194/npg-25-457-2018}

\bibitem[{{Sironi} {et~al.}(2021){Sironi}, {Rowan}, \& {Narayan}}]{Sironi+2021}
{Sironi}, L., {Rowan}, M.~E., \& {Narayan}, R. 2021, \apjl, 907, L44, \dodoi{10.3847/2041-8213/abd9bc}

\bibitem[{{Sironi} \& {Spitkovsky}(2014)}]{SironiSpitkovsky2014}
{Sironi}, L., \& {Spitkovsky}, A. 2014, \apjl, 783, L21, \dodoi{10.1088/2041-8205/783/1/L21}

\bibitem[{{Sobacchi} {et~al.}(2021){Sobacchi}, {Sironi}, \& {Beloborodov}}]{Sobacchi+2021}
{Sobacchi}, E., {Sironi}, L., \& {Beloborodov}, A.~M. 2021, \mnras, 506, 38, \dodoi{10.1093/mnras/stab1702}

\bibitem[{{Stawarz} \& {Petrosian}(2008)}]{Stawarz2008}
{Stawarz}, {\L}., \& {Petrosian}, V. 2008, \apj, 681, 1725, \dodoi{10.1086/588813}

\bibitem[{{Stepney} \& {Guilbert}(1983)}]{StepneyGuilbert1983}
{Stepney}, S., \& {Guilbert}, P.~W. 1983, \mnras, 204, 1269, \dodoi{10.1093/mnras/204.4.1269}

\bibitem[{{Sun} \& {Bai}(2021)}]{SunBai2021}
{Sun}, X., \& {Bai}, X.-N. 2021, \mnras, 506, 1128, \dodoi{10.1093/mnras/stab1643}

\bibitem[{{Tchekhovskoy} {et~al.}(2011){Tchekhovskoy}, {Narayan}, \& {McKinney}}]{Tchekhovskoy2011}
{Tchekhovskoy}, A., {Narayan}, R., \& {McKinney}, J.~C. 2011, \mnras, 418, L79, \dodoi{10.1111/j.1745-3933.2011.01147.x}

\bibitem[{{Walker} {et~al.}(2018){Walker}, {Hardee}, {Davies}, {Ly}, \& {Junor}}]{Walker2018}
{Walker}, R.~C., {Hardee}, P.~E., {Davies}, F.~B., {Ly}, C., \& {Junor}, W. 2018, \apj, 855, 128, \dodoi{10.3847/1538-4357/aaafcc}

\bibitem[{{Werner} {et~al.}(2016){Werner}, {Uzdensky}, {Cerutti}, {Nalewajko}, \& {Begelman}}]{Werner+2016}
{Werner}, G.~R., {Uzdensky}, D.~A., {Cerutti}, B., {Nalewajko}, K., \& {Begelman}, M.~C. 2016, \apjl, 816, L8, \dodoi{10.3847/2041-8205/816/1/L8}

\bibitem[{{Xie} \& {Zdziarski}(2019)}]{XieZdziarski2019}
{Xie}, F.-G., \& {Zdziarski}, A.~A. 2019, \apj, 887, 167, \dodoi{10.3847/1538-4357/ab5848}

\bibitem[{{Xu} \& {Lazarian}(2023)}]{Xu2023}
{Xu}, S., \& {Lazarian}, A. 2023, \apj, 942, 21, \dodoi{10.3847/1538-4357/aca32c}

\bibitem[{{Yuan} \& {Narayan}(2014)}]{Yuan&Narayan2014}
{Yuan}, F., \& {Narayan}, R. 2014, \araa, 52, 529, \dodoi{10.1146/annurev-astro-082812-141003}

\bibitem[{{Zamaninasab} {et~al.}(2014){Zamaninasab}, {Clausen-Brown}, {Savolainen}, \& {Tchekhovskoy}}]{Zamaninasab2014}
{Zamaninasab}, M., {Clausen-Brown}, E., {Savolainen}, T., \& {Tchekhovskoy}, A. 2014, \nat, 510, 126, \dodoi{10.1038/nature13399}

\end{thebibliography}
\bibliographystyle{aasjournal}



\end{document}